# Recent Advances in Tunable Metasurfaces: Materials, Design and Applications

Omar A. M. Abdelraouf [1,3#], Ziyu Wang[3#], Hailong Liu[3], Zhaogang Dong[3], Qian Wang[3], Ming Ye[2], Xiao Renshaw Wang[1,2,*], Qi Jie Wang[1,2,*], Hong Liu[3,*]

**Affiliations:**
[1]School of Physical and Mathematical Sciences, Nanyang Technological University, Singapore 637371, Singapore
[2]School of Electrical and Electronic Engineering, 50 Nanyang Avenue, Nanyang Technological University, Singapore 639798, Singapore
[3]Institute of Materials Research and Engineering, Agency for Science, Technology and Research (A*STAR), 2 Fusionopolis Way, Singapore 138634, Singapore

[#] Equal contribution
*To whom correspondence should be addressed. Email: renshaw@ntu.edu.sg, qjwang@ntu.edu.sg, h-liu@imre.a-star.edu.sg

## ABSTRACT

Metasurfaces, a two-dimensional (2D) form of metamaterials constituted by planar meta-atoms，exhibit exotic abilities to freely tailor electromagnetic (EM) waves. Over the past decade, tunable metasurfaces have come to the frontier in the field of nanophotonics, with tremendous effort focused on developing and integrating various active materials into metasurfaces. As a result, tunable/reconfigurable metasurfaces with multi-functionalities triggered by various external stimuli have been successfully demonstrated, openings a new avenue to dynamically manipulate and control EM waves for photonic applications in demand. In this review, we first brief the progress of tunable metasurfaces development in the last decade and highlight representative works from the perspectives of active materials development, design methodologies and application-driven exploration. Then, we elaborate on the active tuning mechanisms and relevant active materials. Next, we discuss recent achievements in theory as well as machine learning (ML) assisted design methodologies to sustain the development of this field. After that, we summarize and describe typical application areas of the tunable metasurfaces. We conclude this review by analyzing existing challenges and presenting our perspectives on future directions and opportunities in this vibrant and fast-developing field.

## INTRODUCTION

Since the introduction of the concept of abrupt phase changes *via* metasurfaces by Yu *et al.* in 2011,[1] the field of metasurfaces has witnessed a globally rapid development to address the needs for various applications. The huge research effort has led to significant outcomes from the perspective towards tunable metasurfaces, which have been extensively investigated in three typical schemes (Fig. 1), namely design strategies (red box), active materials and tuning mechanisms (blue box) and applications (green box). Geometrical design is the cornerstone of metasurfaces development; therefore, the development of accurate, cost-effective and artificial intelligence (AI) driven design approaches has been propelled to realize on-demand performance superior to conventional optics. Typical design strategies are shown in the red box in Fig. 1. Early work with design strategy inherited from optimization technique was reported to design an acoustic metasurface by using a topology optimization method with the help of finite element method (FEM) till reaching the optimum condition.[2] Furthermore, a more efficient array-level inverse design method has been demonstrated to realize complex functions.[3] With the rise of artificial intelligence (AI), machine learning (ML) has emerged as a promising method for metasurface design by building a training set of fabricated or simulated structures for phase gradient metasurface.[4, 5] These design methodologies have ushered in an era of designer metasurfaces for on-demand optical control.

Recently, much effort has been directed to study active materials, whose optical properties can be controlled under external stimuli such as pressure, heat, current and light. Active materials endow dynamic control of metasurfaces for multi-functionality through the analytical designs of interfacial phase discontinuities by varying the geometrical parameters of meta-elements.[6-8] In this regard, the integration of active materials has been demonstrated to dynamically control light-matter interactions upon external stimuli. Among them, optically induced approaches were heavily investigated as highlighted in the blue boxes of Fig. 1. Ren *et al.* have demonstrated active control of electromagnetically-induced transparency (EIT) in THz communication by integration of photoconductive silicon (Si) islands into Au metasurface unit cells.[9] The increment of photoconductivity of Si under irradiation led to the reduced destructive interference between the resonance modes of two different components of the unit cells. Therefore, the as-fabricated EIT metasurfaces showed a transmission modulation from ~85% to ~50%, as photoexcitation power was increased from 0 to 1000 mW in the THz range. This work offered an emerging solution of using dielectric metasurfaces to develop reconfigurable devices upon light irradiation. Under the dielectric material scheme, both electric and magnetic responses of high refractive index Si nanoparticles with a diameter of ~210 nm could be manipulated under femtosecond (fs) laser irradiation to support magnetic Mie-type resonance.[10] This high-intensity fs laser pulse significantly enhanced optically-induced change of the permittivity of the Si nanoparticles, thereby enabling 20% modulation of the reflection from dielectric metasurface with a laser fluence of up to ~30 mJ/cm$^2$.[11] The pump fluence could be further reduced to ~400 μJ/cm$^2$ to acquire 35% modulation of reflection using direct bandgap GaAs metasurfaces.[12] Subsequently, a milestone work of optically-tunable dielectric metasurfaces using a phase change material (PCM) was realized by Wang $et\ al$..[13] Thanks to the reversible laser-induced phase transition in Ge$_2$Sb$_2$Te$_5$ (GST), the large change of refractive-index could be reverted in the near infrared spectral range (NIR) with different fs-laser pulse intensities.[14] As-fabricated multiplexing optical lens revealed its restored focusing property during the writing, erasing and rewriting processes under the illumination energy of 0.39 nJ, 1.25 nJ and 0.39 nJ, respectively. This technology provided a versatile platform to design and realize reconfigurable multi-focus optics, super-oscillatory lenses and grayscale holograms, which has spurred the study of PCM metasurfaces. Later, transparent conducting oxides (TCO) such as indium tin oxide (ITO) and nanorod metasurface was proposed to control light-matter literation under intraband pumping spanning the NIR to mid-infrared (MIR) spectral range where ITO could retain its metallic phase.[15, 16] These works imply that the tunability of active materials is a pre-requisition for achieving reconfigurable metasurfaces. Similar to the spectral shift, the change of polarization state of light is of equal importance for detection. It was reported that hyperbolic metamaterials with anisotropic permittivity tensors would be rendered transparent and metallic accordingly under polarized illumination with its hyperbolic dispersion regimes in the visible spectral range.[17] Au nanorod arrays in Al$_2$O$_3$ matrix acting as hyperbolic metamaterials have been used to control transmitted light with polarization rotation reaching up to 60$^0$ within sub-picosecond (ps) in the visible frequency.[18] Upon light excitation, the extraordinary ($e$) wave resonances were red-shifted, giving rise to a phase shift of the transmitted $e$-waves while the phase shift of ordinary ($o$) waves remained unchanged. This elaborated the polarization state of the transmitted light could be actively tuned with light control. Cong $et\ al$. further expanded the modulation of polarization state in the hybrid Au-Si metasurface resonators in THz operation range by introducing anisotropic resonance behaviors.[19] These achievements of optically tunable metasurfaces have suggested that ultrafast and high-efficiency active optical devices are promising for applications of imaging, sensing and modulations in the future.

Electrical stimuli are another essential method for achieving dynamic tuning as shown in blue box in Fig. 1. Insulator-to-metal transition (IMT) in metal oxides has manifested their intriguing responses upon different external stimuli such as light, electrical field and magnetic field.[20-22] Utilizing the electrical stimuli, Liu *et al.* have triggered the IMT in vanadium dioxide (VO₂) via a THz electric field, and demonstrated to the transmission variation using arrays of gold split-ring resonators (SRRs).[23] Under high-field THz pump (~1 MV/cm), activation barrier of carrier movements was decreased as a result of Poole-Frenkel (PF) electron release, leading to the acceleration of carriers and subsequent Joule heating. This elucidates that temperature rise yields the IMT of the VO₂, enabling the dynamic response by a THz electrical field. Similarly, Yao *et al.* have incorporated monolayer graphene in a metasurface absorber as a demonstration of the electrically reconfigurable ultrathin optics.[24] On the basis of the broadband tunability of electrostatic doping of graphene and enhanced light-graphene interaction, the graphene conductivity in prefect absorption conditions was around two orders of magnitude smaller than that of pure metasurfaces.[25] Their research has explored the unprecedented opportunities for integration of 2D materials in ultrathin meta-optics.[26] Considerable efforts have been devoted to developing tunable metasurfaces at complementary metal–oxide–semiconductor (CMOS) platform and very recently, Park *et al.* designed an all-solid-state active metasurface that consisted of electrically tunable channels. Each channel was made of a stacked structure of gold (Au) nanoantenna/top insulator/ITO/bottom insulator/ aluminum (Al) mirror.[27] Such design enabled the control of the reflected light and beam steering by applying an electrical bias between the top Au nanoantenna and the bottom Al mirror. A practical light detection and ranging (LiDAR) application has been proposed to perform the depth scan of a street scene, which opens up the commercial opportunity for electrically tunable metasurfaces to be used in well-established CMOS products.

Besides optical and electrical stimuli, other stimuli such as mechanical, thermal, chemical and magnetic inputs, have also been investigated widely, as highlighted in the blue box in Fig. 1. Geometric pattern in the metasurfaces is critical for a variety of optical devices. Stretchable deformation of flexible metasurfaces substrates is a very straightforward method to alter the light-matter integration. However, fabrication of the metasurface structures on flexible substrates remains challenging. To overcome it, Ee *et al*. has developed a metal-assisted transfer method to produce different nanostructures with good retention on polydimethylsiloxane (PDMS) substrates.[28] Thermal stimuli offer another method to introduce dynamic tuning. Park J. *et al*. have demonstrated a typical dynamic control of thermal heating using InAs metasurfaces.[29] Discussions on these approaches will be detailed in the subsequent sections.

The progress of application-driven research is highlighted in the green boxes of Fig. 1. In 2011, it was reported that plasmon-mediated two-photon absorption of nanostructured Au film gave rise to enhanced plasmonic resonance at the wavelength of 890 nm, which showed ~40 fs nonlinear response for ultrafast all optical switching at THz bandwidth.[30] Both convex and concave imaging were achieved by a metalens under visible light, which exhibited dual-polarity functionality as a result of circular polarization-dependent phase discontinuity in 2012.[31] After that, second harmonic generation (SHG) in the NIR was developed from nonlinear metasurfaces on multi-quantum well (MQW) semiconductors.[32] These phase gradient metasurfaces opened up the opportunities for many applications such as metalens,[33, 34] beam steering,[35] hologram,[36, 37] and vortex generation.[38]

Although there are several review articles about tunable metasurfaces in recent years[39-42], in view of its rapid development, we believe a comprehensive review of the latest achievement on active materials, design methods, and applications of tunable metasurface will be beneficial to the community. In this article, we briefly review the progress and achievement in the last decade (2011 to 2021). It starts with a brief introduction with a focus on active materials development, design methodologies and applications. Next, we detail the discussion on the development of active materials, which can be categorized into three groups based on their external stimuli, *i.e.,* electrically tunable materials, optically tunable materials, and materials under other stimuli. After that, we discuss the physics behind tunable metasurfaces, such as Mie resonance, bound states in the continuum (BIC), anapole, and EIT. In addition, with the rapid development of AI, machine learning has been adopted to expedite the design and discovery of tunable metasurfaces. We will analyze and discuss the recently reported machine learning assisted design methods such as particle swarm optimization (PSO), genetic algorithms (GA), neural network (NN), *etc*. Subsequently, we summarize the major applications such as dynamic beam focusing, wavefront shaping and beam steering, and image display. Last but not least, we conclude this review by presenting our perspectives on future directions and opportunities for the next generation of tunable metasurfaces.

**Figure 1. The progress of representative tunable metasurfaces development from 2011 to 2021 on active materials, design and tuning mechanisms, and applications**. Reprinted with permission from AAAS: Science[1] (Copyright 2011) and Science Advance[29] (Copyright 2018). Reference[30] was reprinted with permission from Wiley-VCH Verlag GmbH & Co. KGaA, Weinheim, Copyright 2011. Reprinted with permission from Springer Nature: Nature Photonics[13, 16, 18] (Copyright 2015, 2016 and 2017) Nature Communications[5, 9, 12, 31] (Copyright 2012, 2017 and 2019), Nature[23, 32] (Copyright 2012 and 2014), Nature Nanotechnology[27] (Copyright 2021) Light: Science & Applications.[19] (Copyright 2018). Reference[2] was reprinted with permission from Elsevier B.V. Copyright 2013. Reprinted with permission from American Chemical Society: Nano letters[10, 24, 28] (Copyright 2014, 2015 and 2016) and ACS Nano[3] (Copyright 2020).

## ACTIVE MATERIALS AND TUNING MECHANISMS

**Electrically Tunable Materials.** In this section, we will highlight the recent progress of electrically tunable materials and discuss their mechanisms to realize dynamic tunability. We classify them into four groups, *i.e.*, field effect in TCOs and semiconductors, molecules orientation in liquid crystals (LCs), Pockels effect in ferroelectrics (FEs), and electro-thermal effect in PCMs. Figure 2 summarizes several typical works based on these electrically tunable materials.

**Field Effect in TCOs and Semiconductors.** TCOs are optically transparent materials with a low extinction coefficient. The relative permittivity of the TCOs and semiconductors can be modeled using the Drude model where real and imaginary parts of permittivity depend on plasma frequency. The plasma frequency is defined as the frequency at which the real part of the permittivity approaches nearly zero. Plasma frequency depends on the free carrier concentration in the material. Therefore, the relative permittivity of the material can be altered by introducing modulation of the free carrier concentration.

In this regard, Feigenbaum *et al*. reported an approach to electrically control carrier's concentration in ITO through an external gate or electric fields.[43] Applying positive bias led to the formation of an accumulation layer consisting of the negatively charged carriers. This accumulation surface exhibited different optical properties compared to bulk films, showing that optical properties could be tailored *via* field effect. Shirmanesh *et al*. developed a dual-gate metasurface using an ITO layer with 5-nm thickness to demonstrate dynamic beam steering and a focusing meta-mirror shortly afterwards.[44] This approach was able to offer simultaneous modulation to both amplitude and phase shift of reflected light by varying the applied bias. However, it required an extra step to compensate for this deviation to keep the reflectance constant.

In order to resolve the issue of simultaneous modulation of amplitude and phase shift, Park *et al.* have demonstrated an independent control of phase and amplitude in dual-gated metasurfaces, as shown in Fig. 2a.[27] Once being applied with two different biases, accumulation and depletion layers in ITO could be individually regulated on both electrode interfaces, thus leading to independent control of phase and amplitude of reflected light. Upon variation of the applied voltage, the phase would undergo a complete $\sim 2\pi$ change while the reflectance remained constant. It was suitable for dynamic beam steering and LiDAR but with a relatively low reflected amplitude of < 4%. Lee *et al.* have proposed a relatively high efficiency design of transmission mode using a single-gated metasurface consisting of an ITO layer of 20 nm, reaching up to 30% of transmission efficiency at the wavelength 1550 nm.[45] This work is able to address the need for high transmission efficiency devices in the NIR regime.

Gate field effects have also been investigated in other TCOs and semiconductors for various applications. Aluminum doped zinc oxide (AZO) has been used in gate tunable metasurface in reflection mode.[46] Using atomic layer deposition technique, a layer of AZO was conformally deposited onto germanium nanopillars to increase the surface area of the accumulation layer, and hence it enabled large electrical tunability with a high differential reflection of 40%. Park *et al.* have used an indium arsenide (InAs) layer of 50 nm in a single-gated structure (Fig. 2b), which achieved electrically tuning of the thermal emission of the plasmonic metasurfaces at the MIR wavelength of ~7.2 μm.[29] It is noteworthy that the low carrier concentration of a high-doped ($n^{++}$) InAs layer ($n = 10^{19}$ cm$^{-3}$) resulted in reduced plasma frequency. Similarly, Iyer *et al.* have utilized intrinsic InAs with lower carrier concentration ($n=4.5 \times 10^{17}$ cm$^{-3}$) in the electrically gated tunable metasurface to realize a phase shift of $\sim \pi$ at the wavelength of 13 μm.[47] In contrast, indium silicon oxide (ISO) has a higher carrier concentration ($n = 3.2 \times 10^{20}$ cm$^{-3}$), which yielded a maximum change in the reflection of ~57% at the wavelength of 2.73 μm with a similar gated structure.[48] A *n*-type silicon nano-cone metasurface has been developed to acquire a relatively wide regulation of the IR transmittance, which could be electrically modulated up to 65% at an applied voltage of 5 V.[49] To sum up, field effect offers high modulation depth due to the fast movement of carriers during accumulation and depletion processes, feasibility for CMOS integration, and large change in optical properties.

**Liquid Crystals (LCs)** consist of molecules oriented in different directions which can be controlled by external stimuli, such as an electric field and thermal heating. In general, LCs have four main phases: (1) isotropic phase, where molecules are randomly oriented; (2) nematic phase, where molecules are organized in one direction with no periodicity; (3) smectic phase, where molecules are periodically arranged in one direction between its layers; and (4) chiral phase, where molecules are parallel without periodicity and arranged in a helical shape. The ordered orientation of nematic, smectic, and chiral phases contributes to the anisotropic optical property of LCs. The optical birefringence ($\Delta n = n_e - n_o$) between ordinary refractive index ($n_o$) and extraordinary refractive index ($n_e$) of LCs could reach up to ~0.2.[35] This large contrast of refractive index and the intrinsic transparency of LCs in the visible range suffice for active metasurfaces.[50]

For example, LCs have been integrated with plasmonic metasurfaces to tune their resonances. Driencourt *et al.* combined LCs with silver (Ag) gratings to realize an electrically tunable metasurface acting as a multicolored filter, as depicted in Fig. 2c. LCs were sandwiched between vertical electrodes, yielding 70% of the standard sRGB colors with an applied voltage of 6.5 V only.[51] Kowerdzie et al. fabricated lateral Au electrodes with Au nanocylinders on top of LCs to adjust the plasmonic resonance in the visible spectrum by more than 100 nm under a bias of 13 V.[52] This concept was extended to THz spectrum using two copper split ring vertical electrodes to modulate the transmission amplitude with a modulation depth of up to 96% upon an applied voltage of 16 V.[53] Moreover, the modulated angle of the THz beams was achievable through the LCs embedded between two Au electrodes using field programmable gate arrays (FPGA). Applying an electric bias between individual unit cells, a beam steering angle of $32^{\circ}$ at a frequency of 683 GHz was demonstrated.[54] Similarly, Vasić et al. have proposed a plasmonic metasurface structure with the LCs operating in THz range with an adjustable number of the Au electrodes in each unit cell, which was able to acquire the beam steering angle of up to $75.6^{\circ}$ for the first order of diffraction at an applied voltage of 10 V.[55] Buchnev et al. have replaced both top and bottom metallic electrodes with split ring structures across the LCs to induce phase shift for incident beam at a frequency of 0.8 THz with the measured wavefront steering angle of $4.5^{\circ}$ under an electrical field of 20 V.[56] In addition, typical metallic electrodes are relatively lossy and opaque. Low loss dielectric metasurface and highly transparent electrodes have been strategically developed to alleviate the loss associated with metallic electrodes for the LCs molecular alignment in the visible spectrum. Zhou et al. have proposed a dielectric metasurface consisting of amorphous silicon (a-Si) nanopillars embedded in LCs with two vertical ITO electrodes for switchable metasurface display. Its transmission modulation depth was ~ 53% in visible spectrum under an applied voltage of 20 V.[57] The modulation depth was further increased to 65% using low loss dielectric of $TiO_2$ nanopillars instead of a-Si under an applied voltage of 5 V only.[58] Switchable beam steering was achieved by a combination of the LCs and $TiO_2$ metasurfaces, yielding a tunable steering angle of up to $11^{\circ}$ at an applied voltage of 8 V with a transmission efficiency of 35%, as shown in Fig. 2d.[35] Despite this achievement, better transmission efficiency of up to 60% with an even lower bias has been demonstrated by replacing ITO electrodes with conductive polymers strips.[59] In addition, a highly efficient varifocal metalens with an operation speed at the microsecond level has been illustrated by the structural design of $TiO_2$ unit cells on top of twisted nematic LCs layer in the visible spectrum.[60]

In addition to resonance tuning, LCs have been embedded in dielectric metasurfaces working in the THz spectrum to realize phase shift. Zhang *et al.* fabricated a dielectric metasurface of gradient grating filled with polymer-dispersed LCs. The phase varied from 0 to $\pi$ with a bias variance of up to 80 V at a frequency of 0.8 THz, indicating a tunable THz half wave-plate.[61] Similarly, Ji *et al.* selected silicon rectangular optical elements embedded in the LCs to achieve a polarization conversion ratio of over 96% operating at a frequency of 1.4 THz with a laterally applied alternating current (AC) electric field of a frequency of 90 kHz.[62] Moreover, the metasurfaces based on aluminum gallium arsenide (AlGaAs) nanopillars embedded into LCs matrix can be featured by their ability to reconfigure nonlinear SHG by ~80 nm in transition mode due to the formation of magnetic dipole resonance.[63] Owing to their uniquely modulated optical properties, LCs have been adopted in many applications operating at different spectral regions due to their electrically induced significant refractive index change. Meanwhile, LCs suffer from low switching speed which limits their usage in LiDAR applications and other light modulators applications.

**Pockels Effect in Ferroelectrics.** The Pockels effect is a linear electro-optics effect, which is responsible for the electrical-field dependent optical birefringence ($\Delta n$) of mediums lacking inversion symmetry, such as ferroelectric materials and electrically poled conductive polymers. The relationship between birefringence ($\delta(\Delta n)$) and the electric field ($E$) is governed by $\delta(\Delta n) = \frac{1}{2} r_{eff} n_0^3 E$, where $r_{eff}$ is the effective electro-/optic parameter of materials, $n_0$ is the ordinary refractive index, and $E$ is the applied external electric field. Owing to their optical transparency down to deep ultra-violet (DUV) regime, high modulation speed, stability and inherited large optical nonlinearity, ferroelectric materials have been extensively investigated as one of the promising candidates to constitute active metasurfaces.[64]

Peng *et al.* have theoretically proposed metasurfaces consisting of barium titanate (BTO) single crystal embedded in a 2D grating array with a one-sided microcavity, acting as spatial light modulators (SLM) with a phase shift up to $\pi$ under an applied voltage of 10 V.[65] Such fabricated devices could support continuously tunable beam steering with varifocal length. Karvounis *et al.* later experimentally demonstrated the BTO nanoparticles in the plasmonic metasurface to achieve reflection modulation of up to 0.15% with an applied voltage of 4 V in the NIR as shown in Fig. 2e.[66] This device showed a large operating speed of up to 50 ns and an electro-optic coefficient of

~37 pmV$^{-1}$, which is comparable to lithium niobate (LN) crystals. The LN crystals have also been integrated with metasurfaces to electrically modulate the surface plasmon resonance (SPR). It obtained an efficient intensity modulation of ~5.7 dB/V and a wavelength shift of ~36.3 nm/V in visible band with a low DC current.[67] Similarly, Zhong *et al.* replaced the LN crystals with incipient ferroelectric material KTaO$_3$ (KTO) using top metallic electrodes that supported plasmonic resonance, enabling reconfigurable transmission modulation of ~81% at a frequency of 14 THz and its full width half maximum (FWHM) was increased from 5.46 THz to 8.18 THz with an applied voltage of 60 V.[68] Although ferroelectric materials intrinsically exhibit very high modulation speed, the relatively small change of optical properties somehow limits their applications. It has been foreseen that ferroelectrics have great potential for ultrafast control of optical response with low power consumption, which may find their applications in LiDAR, imaging sensors, camera modules of smartphones, *etc*.

**Electro-thermal Effect in Phase Change Materials (PCM).** In the photonics community, phase change materials (PCMs), typically including germanium antimony tellurium (GST),[13] Ge$_2$Sb$_2$Se$_4$Te (GSST)[69], antimony sulfide family (Sb$_2$S$_3$, Sb$_2$Se$_3$)[70, 71] and VO$_2$, have been intensively investigated. These materials offer a platform for electrically and optically tunable metasurfaces due to their significant refractive index contrast between amorphous and crystalline states, which can be reversibly switched at nanosecond (ns) or even faster speed under electrical pulses (current or voltage) or light excitation. Such electrically tunable optical properties were first applied in a rewritable optical data storage system. With the rapid development of metasurfaces, PCMs have been employed to realize reconfigurable or programmable functionality to manipulate the amplitude, phase, polarization and angular momentum of light. They can be integrated with metallic or dielectric nanostructures to construct hybrid devices or form meta-atoms independently. Besides the amorphous and crystalline states, it has been experimentally validated that GST exhibits intermediate states due to the co-existence of the cubic phase and the hexagonal phase with a different ratio, at which its optical constants can be gradually tuned with proper treatment.[72]

The re-amorphization process of PCMs using electrical pulses is the main obstacle to their application in tunable metasurface. From the engineering point of view, it is challenging to re-amorphize crystalline PCM films as gain mediums, because their high surface-to-volume ratio makes it difficult to apply uniform ultrafast electrical pulses across the whole surface area. In this regard, researchers have developed processes to make nano-antennas using PCMs to fabricate electrically reconfigurable metasurfaces. As shown in Fig. 2f, Wang *et al.* have reported switchable GST/Ag-nanostrip in which the Ag nanostrip acted as both a plasmonic resonator and a heating stage, while the GST nanobeam dynamically tuned the resonances upon pulsed electrical currents.[73] The GST nanobeam could be set by low current pulses with a relatively long period (1 microsecond (µs) duration and 20 µs trailing time) while it could be reset by high current pulses with a short period (500 ns duration and 20 ns trailing time). In addition to the current pulses, another key to switching such low surface-to-volume nanoantenna is that the size of the beneath Ag nanostrip was larger than that of the top GST antenna, allowing the quick dissipation of the heat of melted GST through Ag nanostrip, and thus facilitating GST quenching at the ns-scale. The measured scattering spectra (right panel in Fig. 2f) of the plasmonic resonances corroborated the success of electrically switching of GST nanobeams and good consistency of repeatability. To alleviate the difficulty of the integration of GST nanostrips within functional devices, novel PCMs have been exploited. Recently, Zhang *et al.* have demonstrated dynamically switching of PCM metasurface composed of GSST nanodisks using electrical voltage pulses.[69] In their report, relatively longer and higher voltage pulses (5 µs, 20 V - 23 V) were employed to reset the nanodisks. The reason for employing longer electrical pulses could be attributed to the size and shape of nanodisks and material properties. Moreover, they demonstrated an electrically switchable Huygens' surface beam deflector by individually controlling the phase of nanodisks, which is quite promising for fully integrated, chip-scale reconfigurable optics and may enable potential applications in electrically addressable holograms, lens, beam reflectors, etc.

In contrast to nonvolatile PCMs (*e.g.*, GST, GSST and $Sb_2S_3$), $VO_2$ is a prominent volatile PCM with semiconductor-like properties at room temperature and insulator-to-metal transition under a temperature of > 70 °C, where the crystalline state cannot be maintained without a constant supply of heat. However, its low transition temperature and volatile behavior make it feasible for electrically tuning of phase states. Mohammed *et al.* have demonstrated electrically controlled beam-steering metasurface with $VO_2$ film sandwiched between cross-shaped gold apertures and lightly doped Si substrate.[74] Under an electrical current of 13 mA, a resonance shift of 8 GHz could be realized together with a phase shift of 59°. Furthermore, with the electrical current applied to each unit, phase front would be varied, resulting in beam deflection at a maximum angle of 44°.

In 2017, Zhu *et al*. have placed a VO₂ nanoantenna within the feed gap of a metallic bowtie antenna and demonstrated an electrically modulated VO₂ metasurface with a modulation depth of 33% and a tuning range of up to 360 nm in the NIR region.[75] Dynamically switching the amplitude and wavefront with VO₂ based metasurfaces has enabled potential applications including tunable polarizer and holography.

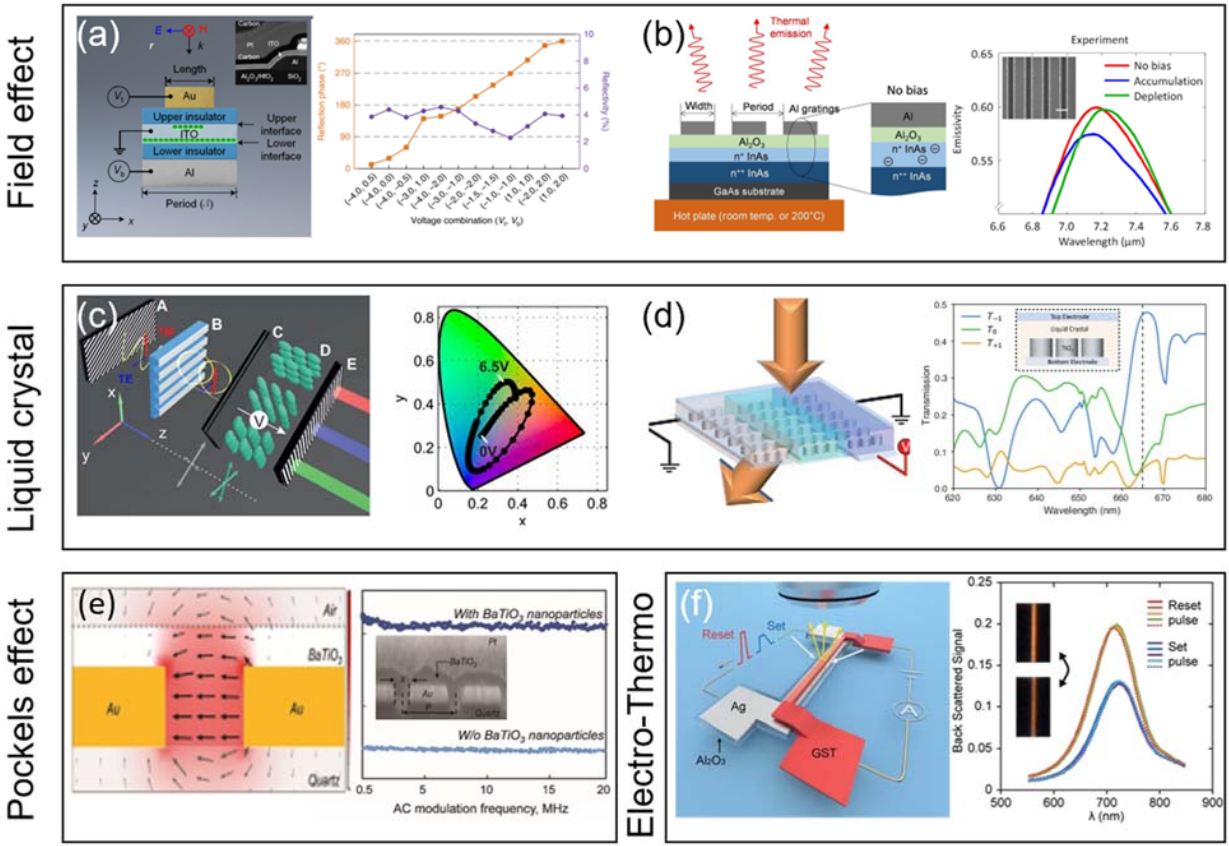

**Figure 2. Electrically tunable materials**. (a) An all-solid-state active metasurface using ITO. Left panel: cross-sectional view of a single plasmonic nanoresonator showing the top Au nanoantenna, the active ITO layer in the middle and Al mirror at the bottom. The upper and lower insulators electrically isolated the three structures from one another, which kept the ITO layer grounded to ensure the top Au nanoantenna and the bottom Al mirror serve as the top ($V_t$) and bottom ($V_b$) electrodes, respectively. Its inset shows a transmission electron microscopy (TEM) image which validated the excellent step coverage on the sloped sidewall of an Al mirror. It indicated that the thin ITO layer was continuously grounded throughout the entire array. Right panel: the measured continuous $2\pi$ phase change with constant reflectivity as a demonstration of the phase-only modulation.[27] Reprinted with permission from Springer Nature, Copyright 2020. (b) A semiconductor plasmonic metasurface for dynamic thermal emission control. Left panel: device schematic. A high-doped (n⁺⁺) InAs layer was epitaxially grown on top of a GaAs substrate and used as a metallic mirror with a negative real dielectric constant at the operating wavelength. A low-doped (n⁺) InAs layer as an active layer offered tunability of its carrier density and concomitant optical properties with an external electrical bias. A thin Al₂O₃ layer serves as both a gate oxide and an optical spacer to a metasurface. The insets show an enlarged schematic for no bias. Right panel: emissivity spectrum for no bias (red curve),

depletion (green curve), and accumulation (blue curve). Its inset shows the SEM image of the fabricated metasurface. Scale bar, $1\mu m$.[29] Reprinted with permission from AAAS, Copyright 2018. (c) A tunable multicolor filter based on LC. Left panel: schematics of elements of the filtering system with (A) an entrance polarizer, (B) the plasmonic nanostructures, (C) a quarter waveplate, (D) a liquid crystal cell and (E) a polarizer. Right panel: CIE color plot of the transmitted colors as a function of the voltage applied to the LC.[51] Reprinted with permission from American Chemical Society, Copyright 2019. (d) Metasurface-based SLM integrated with LC. Left panel: schematic drawing of the SLM and its operation concept. The LC molecules were pre-aligned uniformly along the short axis of the electrodes and parallel to the surface of the substrate. When all electrodes were at the same potential, the LC layer was uniformly aligned in-plane and the incident beam travels straight through the device. When the left/right electrode was biased, the directors of the LC layer on top the electrode were rotated, right/left pointing out-of-plane. The bias difference between the left and right electrodes creates a tilted phase front causing the beam to be deflected to the left/right. Right panel: calculated spectral transmission into the three main diffraction orders: -1, 0 and +1, denoted by $T_{-1}$, $T_0$ and $T_{+1}$, respectively.[35] Reprinted with permission from AAAS, Copyright 2019. (e) An electro-optic ferro-electric metasurface. Left panel: color map of the amplitude of the static electric field excited on the gap between alternating biased Au nanowires. Arrows denote for the distribution of the field on a cross section of the sample. Shaded region corresponds to air and quartz. Right panel: relative reflectivity changes of the optical signal as a function of the AC modulation frequency for the metasurface. Its inset shows the cross-sectional SEM image of a section of the sample illustrating three gold nanowires on silica covered with BaTiO$_3$ nanoparticles ($P$ = 550 nm, $s$ = 175 nm).[66] Reprinted with permission from Wiley-VCH Verlag GmbH & Co. KGaA, Weinheim, Copyright 2020. (f) An electrically switchable PCM antenna. Left panel: Schematic of optical antenna comprised of GST/Ag-nanostrip, separated by a thin Al$_2$O$_3$ layer. Set (blue) and reset (red) current pulses were sent through the Ag nanostrip heater/antenna to transform a-GST to c-GST and back. Right panel: measured spectra of the scattered light from an antenna after reset and set pulses, under dark-field illumination. The dotted lines represent spectra collected after ten cycles of operation. The inset shows two images of the antenna under dark-field illumination after a reset pulse (top) and a set pulse (bottom).[73] Reprinted with permission from Springer Nature, Copyright 2021.

## Optically Tunable Materials.
Ultrafast switching speed is one of the key specifications of the active metasurfaces required for applications of high-speed spectroscopy, communication and LiDAR technologies. Hitherto, it has remained challenging to achieve such ultrafast operation *via* electrical, thermal and electro-mechanical tuning mechanisms. In contrast, optical tuning stands out owing to its intrinsic high speed of light-matter interactions through different approaches such as photocarrier excitation and material phase transition induced by optothermal effect and optical nonlinearities in the constituent materials. Figure 3 summarizes some typical works based on optically tunable materials.

**Photocarriers Generation.** Modulating the free-carrier density and the associated optical properties (*e.g.*, complex refractive index) in conductive materials is a common approach for active metasurfaces. Free-carrier density modulation can be realized by photocarrier excitation under either electrical or optical pumping. Although electrical gating has often been employed for free-carrier injection in conductive materials, the carrier injection and associated optical change generally occur in a nanometer (nm)-thin layer of carrier accumulation due to the field screening effect. In addition, its tuning speed is limited by the electrical resistor-capacitor (RC)-constant of the device. In comparison, photocarrier excitation provides opportunities for ultrafast modulation as its speed is generally dependent on the carrier relaxation time, which is on the order of several or tens of picoseconds. In addition, photocarrier excitation takes place across the whole volume of the materials which offers the possibility to increase the optical modulation depth. Early demonstrations of active metasurfaces were based on photocarrier excitations utilized TCOs[16] and semiconductor materials[9, 10, 19, 76-81]. Being optically transparent and featured with lower free-carrier densities than noble metals, TCOs are favourable materials for the construction of tunable metasurfaces due to their low optical losses, giving rise to effective modulation of the refractive index resulting from the substantial increase in free-carrier density upon photoexcitation. For example, Guo *et al*. demonstrated ultrafast switching of infrared plasmons at sub-picosecond time scale by optically pumping of indium tin oxide nanorod arrays (ITO-NRAs).[16] This was achieved by intra-band photocarrier excitation in the ITO-NRAs and the resultant modulation of its plasma frequency.

Semiconductors, such as Si,[9, 10, 19, 76, 77] germanium (Ge)[81] and gallium arsenide (GaAs),[78-80] provide another choice of material system for optical modulation of charge carrier density. Demonstrations include the use of semiconductor materials as either the surrounding materials[9, 19, 76, 78, 81] or the constituent materials[10, 12, 77, 79, 80] of metasurface structures, whose optical resonances (plasmonic- or Mie-resonances) were tuned by the excitation of photocarriers in the materials and the resultant change of permittivity.

Two-dimensional (2D) materials have also been investigated for optical ultrafast tuning. For example, liquid-exfoliated black phosphorous nanosheets have been reported to exhibit an ultrafast (0.36 ps-1.36 ps) nonlinear optical excitation and saturable absorption in the near- and mid-infrared regions.[82] Picosecond modulation of interlayer Van der Waals (VDW) interactions between quasi-2D transition metal dichalcogenides (TMDCs) have also been demonstrated *via* the above-gap optical excitation.[83] More recently, solution-processed metal-halide compounds[84] and topological insulators[85] have been validated thanks to their ultrafast carrier dynamics and highly sensitive photo-responsive properties. Figure 3a shows an example in which the ultrafast photocarrier dynamics and pump fluence-dependent photoconductivity of solution-processed lead iodide ($PbI_2$) were investigated and a tuning time of less than 150 ps was achieved.[84] The free-carrier excitation and their recombination dynamics in a 160 nm thin film of $PbI_2$ were first studied, which revealed an ultrafast free-carrier relaxation dynamics with a fast relaxation time constant of 8.2 ps -12.8 ps and a slow relaxation time constant of around 50 ps. The study suggested that the fluence dependent fast relaxation time constant of the $PbI_2$ thin film was due to trap-assisted and Auger recombination processes, while the slow relaxation process was related to the electron-hole recombination from trap states to the valence band maxima. The solution-processed $PbI_2$ thin film was then spin-coated onto a THz metasurface which consisted of asymmetric split ring resonators on a quartz substrate, as illustrated in the left panel of Fig. 3a. The transmission spectrum of the $PbI_2$-coated metasurface was featured by an asymmetric Fano resonance at a low frequency (~0.7 THz) and a symmetric broad dipole resonance at a high frequency (~1 THz), referring to the black curve in the right panel of Fig. 3a. Upon optical pumping with a photon energy of 3.09 eV (larger than the bandgap of the $PbI_2$ thin film), photoexcited free carriers in the $PbI_2$ film within the resonator gaps strongly altered the metasurface resonances and led to a substantial modulation of transmission. The modulation speed of less than 150 ps was achieved together with modulation depths of unity and 50% at the Fano and dipole resonant frequencies, respectively (Fig. 3a, right panel).

Figure 3b shows another work which employed topological insulator $Bi_2Se_3$ as the photoactive material in a hybrid EIT metasurface.[85] This active $Bi_2Se_3$-hybrid EIT metasurface consisted of Π-shaped meta-atoms connected by micropatterned $Bi_2Se_3$ bridges (grey strips, ~20 nm thick) epitaxially grown on sapphire substrate, as schematically illustrated in the left panel of Fig.3b. Due to the destructive interference between an electric bright mode and a magnetic dark mode, an EIT window was induced at a frequency of ~0.67 THz. Under optical pumping, photocarriers were excited in the $Bi_2Se_3$ bridges, which resulted in a short-circuit of the dark resonance mode and therefore switching-off to the EIT. Consequently, a transmission modulation of up to 31% within ~9.5 ps was observed (Fig. 3b, right panel).

**Opto-thermal Effect in PCMs.** In the previous works of PCMs-based reconfigurable metasurfaces, the phase change was activated either by purely thermal heating or electrothermal heating. However, their operation speed normally was slow, due to the slow temperature rising processes driven by direct heating or electrothermal effect. In contrast, opto-thermal effect under ultrafast optical excitation has been validated to be able to complete the phase transition processes within a much short time scale (~ps). Such PCMs include $VO_2$[86] and GST.[23, 87-90] Ultrafast switching and transmission modulation were demonstrated by $VO_2$-embedded hybrid metasurfaces.[90] The device consisted of a square array of two folded gold wires connected with a $VO_2$ pad. Without optical pumping, the structure supported a single bonding dimer plasmonic (BDP) mode. Upon optical pumping ($\lambda$ = 800 nm, laser power of ~1.68 W), a mode splitting occurred due to the significant conductivity change (4-5 orders of magnitude) of the $VO_2$, because it underwent phase transition from an insulating monoclinic phase to a metallic tetragonal phase. As a result, the transmission spectrum was modulated up to around 50% at the original BDP resonance frequency (~0.62 THz) within an ultrashort time of 149 ps. To facilitate the phase transition under low pumping fluences, the sample was pre-heated to 64 ºC which is just below the critical temperature of $VO_2$ (68 ºC). It suggested that the underlying mechanism for the phase transition was due to a combination of the fast Mott-Peierls phase transition and a slow thermally activated grain growth and percolation process.

Besides the electro-thermo effect mentioned before, the state change of GST can be triggered by the opto-thermal effect as well. To re-amorphize GST from crystalline to an amorphous state, the crystal lattice must be molten and subsequently quenched into an amorphous phase at room temperature at a very rapid cooling rate of $10^9$ to $10^{10}$ K/s to avoid any recrystallization of the atomic structures.[91] Such a quench rate can be realized by applying ultrashort laser pulses (or electrical pulses mentioned before) together with a thermally designed structure.[92] Ultrashort ns- and fs-pulses have been adopted for optically switchable GST metasurfaces.[13, 93, 94] In 2020, Aleksandrs *et al.* demonstrated a Huygen's metasurface which was composed of GST/Ge/GST nanodisks. Upon tuning the states of GST, Kerker condition was satisfied, resulting in a high transmission and $2\pi$ phase change. Pixel-by-pixel switching of the phase states of nanodisks with a single laser pulse hide encoding images into Huygen's metasurface. By employing hyperspectral imaging measurements, they directly observed the amplitude and phase distributions of such encoded images. Figure 3c shows the optical tuning of GST metasurface response in terahertz split-ring resonators.[89] An asymmetric split-ring resonator was designed to support sharp Fano resonances at terahertz frequencies which were modulated upon optical pumping of the GST (Fig. 3c, left panel). The Fano resonance was modulated due to the increased THz conductivity of the GST film. The dynamics of the Fano modulation was experimentally studied by varying the pump-probe delay time (Fig. 3c, right panel). The Fano resonance was completely switched off within ~4 ps and fully recovered in ~19 ps, demonstrating the ultrafast tuning capability of GST-integrated metasurfaces by an optical stimulus. Interestingly, it was also demonstrated that the response time can be varied by thermally controlling the phase and crystalline order of the GST film, thus enabling ultrafast tuning with variable tuning speed.

Although GST is the most popular PCM, it exhibits a relatively large extinction coefficient ($k$) in the visible and near-infrared regions, resulting in remarkable optical losses and limitation in the functionality of active metasurfaces. To overcome this problem, low loss GSST[95] and other novel PCM families such as $Sb_2S_3$[70], $Sb_2Se_3$[96] and $As_2S_3$[97], have been investigated recently. For example, four GSST nanodisks with different diameters have been employed as a phase platform, in which 16-level phases have been realized through optically addressing each nanodisk, thereby enabling dynamically tunable optical vortex and holograms. In addition to amorphous and crystalline states, intermediate states can also be reached to realize multi-level complex beam steering and hologram design.[98] The difference of refractive index of $Sb_2S_3$ between amorphous and crystalline states is larger than 1 in the visible region. That makes $Sb_2S_3$ more appropriate for visible photonic structures and metasurfaces. Dong and co-workers demonstrated that $Sb_2S_3$ could be switched among amorphous, intermediate, and crystalline states with electrical and optical tuning methods. Figure.3d shows the optical tuning of $Sb_2S_3$ thin film, which was stacked as $Si_3N_4$/$Sb_2S_3$/$Si_3N_4$/Al layers. Upon increasing the laser power from 2.0 mW to 10.0 mW, the $Sb_2S_3$ film was gradually tuned from an amorphous state to an intermediate state and then to a crystalline state. With the gradually changed refractive index, gradient colors were achieved. Rewritable color printing has been demonstrated with a pixel-by-pixel optically addressable method.[70] To sum up, PCMs can offer a rich variety of unique merits such as strong modulation of complex refractive index (typically $\Delta n > 1$) upon phase transition, reversible switching across multiple crystallographic phases by optical, electrical and thermal means, low switching energy[99], sub-nanosecond switching speed[100] and high cyclability[101] of up to $10^{15}$. Thus, they can enable reconfigurable photonics for various applications mentioned above.

**Opto-thermo Effect in Other Materials**. The opto-thermo effect occurs when bulk materials absorb external thermal energy, which changes their optical properties, volume and lattice structure. The opto-thermo coefficient (OTC) is defined as the change in refractive index with the response to temperature, which can be either positive or negative depending on the material physical properties. Bosch *et al.* have fabricated a thermally tunable polarization converter using germanium nanopillars operating in NIR.[102] Because germanium has low loss in NIR and its OTC is $\sim 4.5 \times 10^{-4}$ °C$^{-1}$, the corresponding Fano resonance was shifted by more than 40 nm with a temperature change of 100 °C and a transmission amplitude modulation of ~50%. Stephanie *et al.* proposed a theoretical model consisting of silicon nanopillars (its OTC is $\sim 2 \times 10^{-4}$ °C$^{-1}$) to thermally modulate the transmitted wavefront.[103] By creating a high Q-factor Fano resonant metasurface operating at the wavelength of 1521 nm, a significant change in the amplitude and phase of transmitted light was achieved with a temperature change of 250 °C. Free-space modulation and wavefront-shaping modulation based on the nonlocal metasurfaces could be switched on/off at different temperatures.

Strontium titanate (STO) has been employed in thermally tunable metasurfaces due to its temperature-dependent permittivity. Zhong *et al.* had demonstrated quite a few tunable metasurfaces based on STO operating at THz regimes, such as single mode absorbers,[104] multi-mode absorbers[105] and multi-mode light modulator[106] with a temperature change of 90 °C. Similarly, Xin *et al.* have theoretically proposed a perfect absorber consisting of a STO layer with a thickness of 2 μm and plasmonic Au nanostructures, which was able to work from 1.71 THz to 2.48 THz with a temperature change of ~200 °C.[107]

Barium strontium titanate (Ba$_x$Sr$_{1-x}$TiO$_3$, or BST) is a well-known perovskite ferroelectric material, and its permittivity can be changed under an externally applied electric field in the microwave and THz range. Recently, Wu *et al.* have proposed BST as a thermally tunable material after observation of three main phase transitions in its lattice structure, including cubic-tetragonal, tetragonal-orthorhombic, and orthorhombic-rhombohedral transitions upon cooling down to the cryogenic temperature.[108] BST film was integrated with a plasmonic split ring resonator and a transmission modulation of up to 45% at 1 THz under a temperature modulation of 200 °C was achieved.

Last but not least, temperature-responsive polymers can change their volume upon absorbing thermal energy, which can be leveraged to constitute polymer-based thermally controllable metasurfaces. Poly(N-isopropylacrylamide) (PNIPAm) is a typical thermo-responsive polymer and has been included in the metal-insulator-metal Fabry-Perot cavity to adjust the transmitted light in the visible range.[109] Upon applying thermal energy, this PNIPAm layer underwent a swelling effect to reduce its thickness. As a result, a large shift of colors across the visible spectrum under a temperature change of 20ºC was experimentally demonstrated, which shows a potential in future display devices. In addition, Magnozzi *et al.* utilized plasmonic gold nanoparticles coated with the PNIPAm to control the localized surface plasmon resonance (LSPR).[110] The polymer thickness in the shell layer varied from 312 nm at 20 ºC to 162 nm at 50 ºC, leading to LSPR shift by 32 nm. This thermo-responsive mechanism has also been applied in an anionic fluorescent dye (Atto 532). Huang *et al.* added a layer of Atto 532 with a thickness of 10 nm between Ag nanocubes and metallic film to create a plasmonic cavity.[111] Upon being applied with heat, the peak wavelength of plasmon resonance can be tuned over 45 nm in the visible spectrum.

**Optical Nonlinearity.** The excitation of optical nonlinear responses usually requires high-intensity ultrafast optical pumping (*e.g.*, by fs laser), which results in the generation of light signals with frequencies/intensities different from that of the pumping light. Therefore, optical nonlinearity provides a platform for the construction of ultrafast tunable metasurfaces. For example, optical nonlinearity in plasmonic nanostructures has been employed to achieve THz bandwidth all-optical switching of transmission[30] and polarization states at a picosecond level.[18] By utilizing the large optical nonlinear responses in epsilon-near-zero (ENZ) materials, ultrafast tuning of optical nonlinearities has also been demonstrated in ITO[112] and other metal oxides[113-116]. The large nonlinear optical response at the ENZ wavelength, also known as the ENZ effect, describes the phenomenon that occurs over certain spectral ranges, within which the real part of the permittivity vanishes upon a small change in permittivity and produces to a substantial change in the refractive index. Figure. 3e shows high harmonic generation (HHG) using a thin film of low-loss, indium-doped cadmium oxide (In-doped CdO) based on the ENZ effect.[115] In addition, the large field enhancement obtained at the ENZ wavelength due to the Ferrell-Berreman resonance further improved the HHG efficiency by boosting the pumping field intensity.[117, 118] As illustrated in the left panel of Fig. 3e, the sample consisted of a 200-nm-thick Au film coated on a 75-nm-thick In-doped CdO layer epitaxially grown on a (100)-oriented magnesium oxide (MgO) substrate. The In-doped CdO had a measured carrier doping density of $2.8 \times 10^{20}$ cm$^{-3}$ and electron mobility of 300 cm$^2$ V$^{-1}$ s$^{-1}$. The real part of its permittivity crossed zero at a wavelength of 2.1 μm. High harmonic radiation of up to the ninth order was generated by pumping the sample from the substrate side at an angle of 50º with a *p*-polarized 60 fs laser pulse at a repetition rate of 1 kHz centered at 2.08 μm. Notably, the harmonic peaks exhibited red-shifts away from the expected harmonic photon energy, due to the photo-induced heating of electrons in the conduction band and consequent time-dependent ENZ wavelength of the CdO film. This was illustrated by the experimentally determined time evolution of the cavity resonance wavelength in the right panel of Fig. 3e, indicating a fast increasing of cavity resonance wavelength at a sub-picosecond timescale.

To further enhance the nonlinearity response, a metasurface comprising optical dipole antennas coupled to an ENZ material (ITO) was created.[119] The left panel of Fig. 3f shows a schematic of the coupled ENZ-nanoantenna device consisting of a two-dimensional array of gold antennas on a 23-nm thick layer of ITO over a glass substrate. The gold antennas were designed to resonate at the ENZ wavelength of the ITO film, which was around 1420 nm. Under optical pumping, the gold antennas coupled the incident light into the ENZ layer and enhanced the optical field intensity. This resulted in an increment of the nonlinear refractive index for a constant incident intensity compared to the bare ENZ layer without optical antennas and a decrease of the threshold intensity for the onset of the nonlinear response. Consequently, the measured maximum effective nonlinear refractive index ($n_2$) was as large as -3.73±0.56 cm$^2$/GW, which was 2000 times larger than that of a bare ITO. In addition, depending on the spectral position of the pumping light relative to that of the antenna resonance at low pumping intensity, both the sign and magnitude of the effective nonlinear refractive index could be controlled. The temporal dynamics of the nonlinear response was also studied through a pump-probe transmittance measurement at a pumping wavelength of 1280 nm, showing a rise time of ~260 fs and a recovery time of ~600 fs (Fig. 3f, right panel).

It should be noted that large nonlinear optical responses have also been reported in LCs.[120, 121] However, its response time was relatively slow (~milliseconds to several seconds). Recently, nonlinearities in LCs have been explored to demonstrate all-optical switching of a dye-doped LC plasmonic metasurface based on the surface-induced nonlinear effect.[122] The switching time of the reported metasurface was on the order of several seconds, due to the slow re-orientation process of the LC molecules. It was suggested that the switching time of nonlinear responses in LCs could potentially be reduced to milliseconds by using ns-pulse excitation,[121] yet which was still much longer than the response time of tunable metasurfaces based on nonlinearity tunings using other materials discussed in this section.

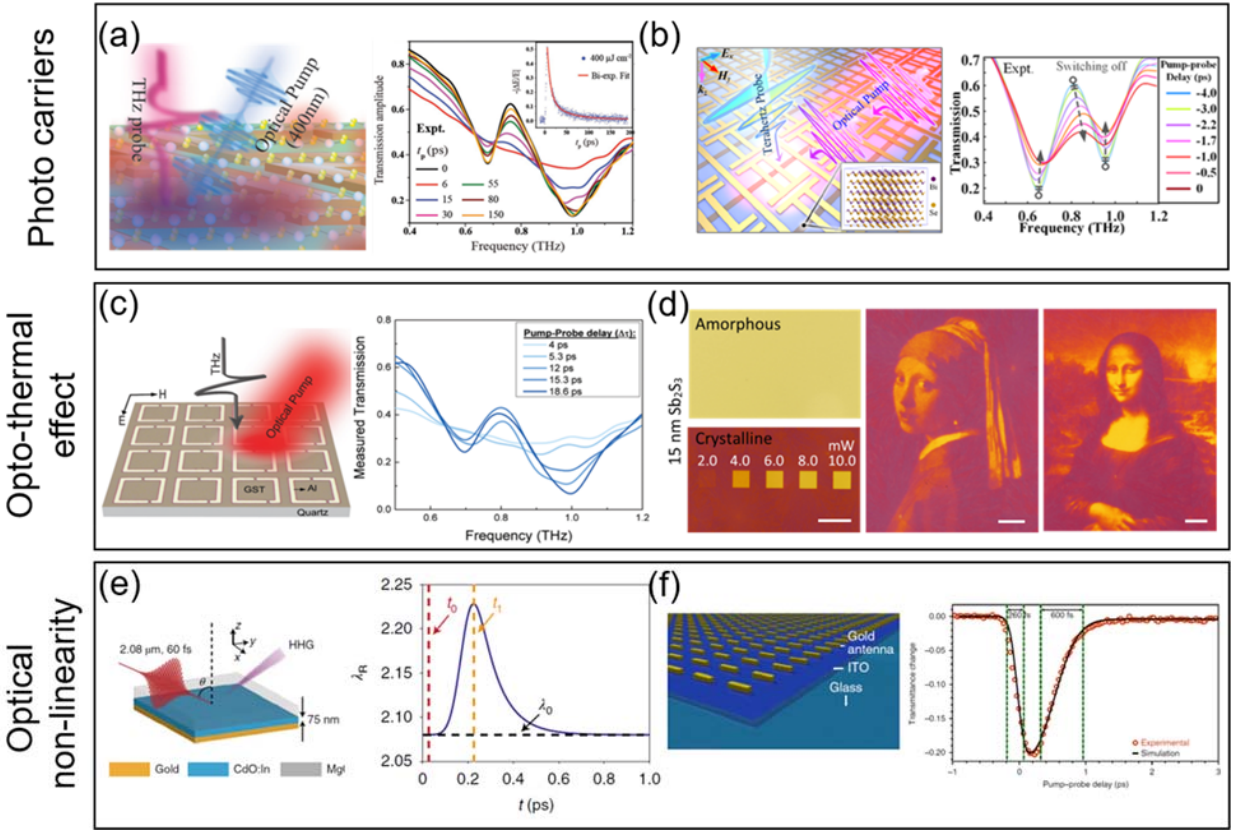

**Figure 3. Optically tunable materials.** (a) Ultrafast all-optical switching terahertz metasurfaces based on photocarrier excitation in solution-processed lead iodide (PbI₂).[84] Left panel: schematic of the switching device consisting of PbI₂ spin-coated on top of split ring resonators. Right panel: modulated transmission spectra measured at various pump probe delays ($t_p$) at a pump fluence of 400 μJcm⁻². Reprinted with permission from Wiley-VCH Verlag GmbH & Co. KGaA, Weinheim, Copyright 2019. (b) Topological insulator Bi₂Se₃-functionalized metasurfaces for ultrafast all-optical modulation of terahertz waves.[85] Left panel: schematic illustration of the active Bi₂Se₃-hybrid metasurface consisting of II-shaped meta-atoms connected by micropatterned Bi₂Se₃ bridges. The THz wave was polarized along the $E_x$-direction and normally impinged on the surface. Inset shows the schematic diagram of the atomic structure of few-layer Bi₂Se₃. Right panel: experimental transmission spectra measured at different pump-probe time delays in the switching off process. Reprinted with permission from American Chemical Society, Copyright 2021. (c) Optically controlled phase transition in GST for ultrafast terahertz wave modulation.[89] Left panel: schematic of the device consisting of resonant asymmetric split-ring resonators integrated with GST. Right panel: ultrafast THz modulation with measured transmission spectra at different pump-probe time delays at a pump fluence of 636.6 μJ cm⁻². Reprinted with permission from Wiley-VCH Verlag GmbH & Co. KGaA, Weinheim, Copyright 2019. (d) Rewritable chalcogenide microprints using antimony sulfide (Sb₂S₃). Left panel: amorphous, crystalline, and optically re-amorphized Sb₂S₃ chalcogenide PCM. Rewritten micrographs of Girl with a Pearl Earring (Middle) and Mona Lisa (Right).[70] Reprinted with permission from AAAS, Copyright 2020. (e) Ultrafast HHG from an In-doped CdO thin film.[115] Left panel: schematic of the ENZ sample composed of (top to bottom), namely gold capping layer, In-doped CdO layer and MgO substrate. The sample was pumped from the substrate side with an incident angle of $\theta$ in free space. Right panel: experimentally determined time evolution of the cavity resonance wavelength within a sub-picosecond timescale under optical pumping. Reprinted with permission from Springer Nature, Copyright 2019. (f) Picosecond tuning of large optical nonlinearity of an ENZ ITO film coupled to plasmonic dipole nanoantennas.[119] Left panel: schematic of the metasurface consisting of a 27-nm thick gold dipole antenna array on a glass substrate. A 23-nm-thick ITO layer was sandwiched between the antenna array and the glass substrate. Right panel: temporal dynamics of the pump-probe induced transmittance modulation at a pumping wavelength of 1280 nm, showing a rise time of ~260 fs and a recovery time of ~600 fs. Reprinted with permission from Springer Nature, Copyright 2018.

**Active materials under Other Tuning Stimuli.** In this section, we will review the recent progress on tuning mechanisms using other stimuli, which are categorized as mechanical, chemical and magnetic stimuli. Figure 4 shows those works on active materials and metasurfaces which can be excited by such external stimuli. They have been validated and adopted as effective approaches to alter the optical performances of metasurfaces and, hence, provide the versatility for practical applications, such as communication, sensing and display.

**Mechanical Stimulus.** The use of elastic substrates integrated with metasurfaces yields effectively controllable optical performances. Stretching flexible photonics makes geometric deformation of nanostructures on elastic substrates, therefore leading to tunable EM field enhancement. Recently, much effort has been devoted to the realization of stretchable metasurfaces onto elastic substrates via a nanofabrication process.[123, 124] Because of polar etching solution (*e.g.* acetone) involved, the flexible substrates, such as PDMS, could be dissolved, resulting in irreparable damage to the nanostructure.[28, 125, 126] In order to solve this issue, Liu *et al*. have developed a metal-assisted transfer method to produce different nanostructures, such as disk array and bull's eye structure on soft PDMS substrates, which were particularly suitable for forming a tiny metal gap.[127] A Ag film was pre-coated on silicon substrates and could be easily peeled off whilst the EBL-processed nanostructures with sub-10 nm gap would be well-retained after being transferred onto targeted elastic substrate slabs. The as-fabricated bowtie structure with sub-10 nm feature is presented in Fig. 4a, and the gap width could be well controlled by applying strain, thereby causing a red shift in backscattering spectra from 810 nm – 927 nm with decreased strain. Moreover, this metal-assisted transfer method presents the potential to assemble diverse 3D nano- and micro-structures from transferred 2D precursors, after being releasing from the soft substrate.

Apart from being capable of tuning scattering property, elastic metasurfaces with highly coupled ring resonators and sufficiently fine gap features exhibit highly anisotropic permittivity.

Hence, it has demonstrated giant phase retardations (up to 180º) of transmitted light in THz, which potentially enabled half-waveplates (HWP) in single-layered metasurfaces.[128] Upon mechanical stretching, the gap would be increased to 108 μm whilst it reduced the strength of the coupled resonators and decreased phase retardation to ~90º, suggesting the functionality of a quarter-waveplate (QWP) as shown in Fig. 4b. Such mechanically tuned resonance coupling has been widely manifested in various stretchable photonic applications.[129, 130] Jia *et. al.* and Gupta *et. al.* have developed the diverse metal plasmonic nanostructures on plastic tapes and inside PDMS slabs, which have both been demonstrated as strain sensors with respect to the strain-induced resonance shift.[131, 132]

In addition, varifocal metalenses were developed under the mechanical deformation of flexible substrates.[28, 133] The focal length could be varied as a result of optical wavefront change due to spatially adjusted plasmonic individual resonators in the stretchable substrates, thereby enabling over 40% modulation of the focal length. Beyond planar metasurfaces, their 3D counterparts could yield enhanced chiroptical response due to an additional dimension perpendicular to the 2D metasurface.[134] As suggested by Liu *et al.*, stereo metasurfaces could readily be produced by mechanical deformation of flexible substrates instead of using complicated nanofabrication process.[127, 134] Owing to stretchable twisted breaking symmetry in three dimensions, the enhanced Fano resonances responded oppositely under the illumination of the left and circularly polarized light, yielding higher circular dichroism intensity than that of their 2D structures. On the other hand, flexible metasurfaces were limited by their anisotropic performance due to inevitably simultaneous stretching and compression at two orthogonal directions. To alleviate this problem, polarization-independent dielectric $TiO_2$ metasurfaces in PDMS have been developed.[135] The as-fabricated metasurfaces with a periodic nanoblock array exhibited consistent tunabiliy under orthogonally polarized reflective light due to systematic effect from grating and near-field mutual interaction. Based on these efforts, deformable metasurfaces have become an emerging technology to introduce highly tunable and efficient photonic applications while retaining the simplicity of the fabrication process.

Integration of tunable metasurface on silicon-on-insulator is of strategic significance in modern integrated optoelectronic systems. The combination of micro-electromechanical system (MEMS) with metasurfaces will offer dynamic reconfigurability for distinct applications such as LiDAR and imaging.[136, 137] Oshita *et al.* have fabricated Au grating structures on MEMS cantilever for tunable plasmonic photodetection.[138] Under an acoustic pressure, the incident light could be adequately tilted from -21.0º to +21.0º, which could reversely match SPR coupling condition in the wavelength range from 1200 nm to 1500 nm. Therefore, reconfigurable SPR signals would be detected as photocurrent which was useful for spectroscopic measurement. Furthermore, as shown in Fig. 4c, it illustrates a suspended metasurface consisting of silicon antenna arrays, each of which could function as an individual Mie resonator for light manipulation. It could continuously modulate the phase (from 0 to $2\pi$) and amplitude of the incidence.[139] With this structure, individual Mie resonance and Fabry-Pérot modes between suspended nanostructures and silicon worked systematically to control the scattering. By applying low actuation voltages from 0 V to 2.75 V, the spacing between suspended metasurfaces and silicon substrate would be varied by hundreds of nm, leading to visible color change. In this regard, temporal color mixing, was demonstrated under different duty cycles with an electrical bias between 0 V and 1 V. More importantly, the width-graded nanobeam structures varied from 80 nm to 160 nm and vice versa. Both concave and convex flat lenses with an electrically tunable focal length could be achieved.

Owing to limited stretching distances, the singlet metasurface lenses are constrained by the tunability of focal length, causing their difficulty in achieving the functionality of compact varifocal lens within a large area.[140, 141] Doublet rotatory flat lenses configurations, such as Alvarez and Moiré concepts, have been demonstrated to mitigate this limitation.[142] By introducing cubic phase elements in the Alvarez lenses, Colburn *et al.* have theoretically and experimentally demonstrated the modified metasurface lenses capable of focusing an Airy beam that could support extended depth of focus (EDOF) with insensitive point spread function (PSF), as depicted in Fig. 4d.[143] More importantly, the fabricated lenses showed dynamic control over EDOF with nearly invariant PSFs from 1.2 mm to 6 mm by varying mutually lateral displacement from 25 μm to 125 μm and tunable EDOF at the entire wavelength range from 440 nm to 640 nm at a fixed displacement, which yielded a potential achromatic focusing under broadband irradiation. Adopting the Moiré scheme, Wei *et al.* have fabricated a polarization-independent metalens doublet based on silicon nanocylinders arranged in a square lattice which exhibited adjustable focal lengths from ±3 mm to ±54 mm by varying the relative angle between the two metalenses while fixing their distance. A zoom factor of an imaging system as large as 18 was realized.[144] Furthermore, a generalized design principle has been proposed for metalens pairs based on silicon nanorods by either lateral or rotational motions to achieve desired functions, therefore paving the application for communication and imaging.[145]

**Chemical Stimulus**. Chemical reactions provide a straightforward method to alter optical properties of materials and enable dynamic optical response while physically maintaining structures.[146] Upon intercalation of Li ions, the reflectance spectra of $WO_3$/Si film reflectors were blue-shifted up to 107 nm and a high transmission contrast was achieved, experimentally validating their potential application for the full color-tunable display.[147] Chen *et al*. have demonstrated highly conductive PEDOT:Sulf polymer nanoantennas supporting plasmonic resonance, which was reversely shifted over 1.3 μm at the oxidized and reduced states (Fig. 4e).[148] In addition, the phase transition via hydrogen uptake in metals, such as magnesium (Mg-$MgH_2$), could yield diverse optical responses, in which dual photonic functionalities (*e.g.* holography and color display) could be realized in a single device with spatially arranged subwavelength pillars.[146] These findings on chemically-induced tunable nanophotonic have paved a way towards the implementation of polymer nanophotonic for compact, multifunctional and cost-effective meta-optics and future display.

**Magnetic Stimulus.** Magnetic stimulus provides uniquely contactless solutions to enhance the versatility of metasurfaces. Complementary design of nano-antennas using $N_{18}Fe_{19}$/Au is illustrated in Fig. 4f, and the optical properties of this material can be changed under an external static magnetic field.[149] Magnetic modulation of refection and transmission spectra at far field and enhancement of magnetically-tuned electrical and magnetic fields by five times at near-field were demonstrated, which was exclusively attributed to magneto refractive effect in giant magnetoresistance material for highly efficient sensing. Wang *et al*. proposed oxide-nitride (NiO) hybrid metasurfaces consisting of the vertically aligned NiO nanorods embedded in titanium nitride (TiN) matrix as a solution to produce metal-free hyperbolic metamaterials.[150] Because of the ferromagnetism of the NiO and surface plasmonic mode supported by TiN matrix, the magneto-optical (MO) coupling and the ferromagnetic response of the NiO/TiN could be actively modulated by an external magnetic field. Upon coupling, two different hyperbolic dispersion regimes (types I and II) were formed at different frequencies. Anisotropic ferromagnetism and enhanced MO effect made the NiO/TiN compound suitable for magneto-optical integrated devices with low optical loss towards highly reconfigurable nanophotonic devices. Nevertheless,

Shamuilov *et al.* have demonstrated generalized concepts for optical magnetic lenses using naturally MO materials (*e.g.*, yttrium iron garnet (YIG)) and 2D materials (*e.g.*, graphene) to achieve tunable focal length in a broad spectral range, which can be used for the design of magneto-optical lens in a wide range of applications.[151]

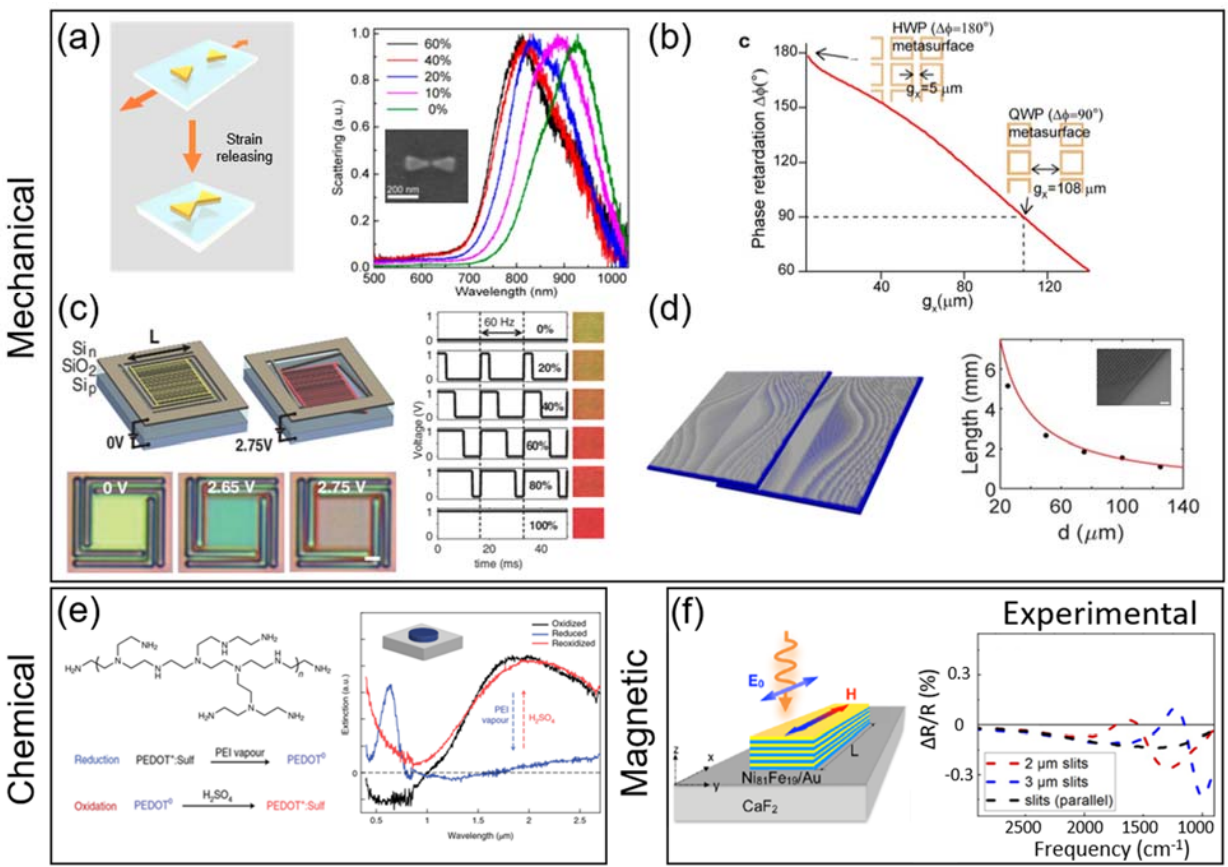

**Figure 4. Active materials under other tuning stimuli**. (a) Mechanical stimulus. Left panel: schematic of a mechanical metasurface on a stretchable substrate. Right panel: measured scattering cross section for different stretching percentages with its inset shows the SEM of a single unit cell.[127] Reprinted with permission from American Chemical Society, Copyright 2018. (b) Mechanically deformable metasurface. Left panel: optical images of fabricated metasurface for unstretched (top) acting as half-wave plate and stretched (bottom) acting as a quarter-wave plate. Right panel: numerical calculation of phase retardation with different gap ($g_x$) spacing.[128] Reprinted with permission from American Chemical Society, Copyright 2019. (c) MEMS-based metasurface. Left panel: electrical actuation with a voltage source (top) and optical images of fabricated metasurface under different applied voltage (bottom). Right panel: measured reflective color images under different duty cycles of applied voltage using fixed frequency ~ 60 Hz.[139] Reprinted with permission from AAAS, Copyright 2019. (d) Tunable EDOF lens. Left panel: schematic of lateral actuation of a pair of wavefront-coded metasurfaces. Right panel depicts the change in focal length in mm for different lateral displacement in $\mu$m between the two metalens (inset: SEM image of a fabricated metalens, Scale bar, 1 $\mu$m).[143] Reprinted with permission from American Chemical Society, Copyright 2019. (e) Chemical stimulus. Left panel: optical properties of conductive polymer can be switched when it changes from metal (PEDOT:Sulf) to dielectric (PEDOT) through a reduction in PEI vapour and reverse switching through oxidation. Right panel: measured extinction cross section of polymer nano-disk (inset) after switching and reverse switching.[148] Reprinted with permission from Springer Nature, Copyright 2019.

(f) Magnetic stimulus. Left panel: schematic of magnetically tunable metasurface consisted of multilayers of magneto-resistance materials $Ni_{81}Fe_{19}$/Au. Right panel: measured normalized differential reflection for parallel light polarization to grating (black), and perpendicular light polarization (red and blue).[149] Reprinted with permission from Optica Publishing Group, Copyright 2020.

## PHYSICS AND DESIGN METHODOLOGIES

**Physics Behind Tunable Metasurfaces.** In this section, we are going to review the recent advances of the new physics to drive the development of tunable metasurfaces (Fig. 5), which include Mie resonances, multipolar decomposition, BIC, anapole and EIT.

**Mie Resonances.** When light interacts with high refractive index spherical nanoparticles with the effective size of being comparable to incidence wavelength, the optical field scattered from the nanoparticles can be solved mathematically, *i.e.,* Mie scattering.[152] High refractive index nanostructures with Mie resonance are generally termed as dielectric nanoantennas, while typical dielectrics consist of Si, $TiO_2$, GaP, GaN, GaAs, and so on. Moreover, metasurface based on dielectric nanoantennas with Mie resonances has been extensively investigated for various applications, such as structural colors,[153-158] optical holograms,[36, 37, 159] field enhancements,[160-165] directionality engineering,[11, 166, 167] harmonic generations[167-171] and so on.

Recently, several investigations have been carried out on tunable dielectric metasurfaces based on Mie resonance. For example, a non-volatile optically controllable metasurface was theoretically investigated at the operating wavelength of 1.55 μm by utilizing the low loss PCM,[172] which is a potential candidate to replace the conventional high-index dielectric materials.[154, 163] Here, the GSST nanobar array was the key building block supporting various Mie resonances, such as magnetic dipole (MD) and electric dipole (ED) resonances, which could be tuned or switched by controlling the crystallization level of GSST. For instance, Figure 5a presents the schematic of the design, where a 300-nm-thick GSST nanobar with a width of 300 nm was placed on a magnesium fluoride ($MgF_2$)/Au substrate. $MgF_2$ was chosen here due to its low dielectric constant of 1.9044.[172] The Au reflecting mirror (200-nm-thick) increased the interaction of the incident optical field with the GSST nanobars, which were periodically arranged along the *x*-direction and being infinitely long along the *y*-axis. The incident optical field was the transverse electrical (TE) polarized plane wave. In addition, Figure 5a presents the simulated near-field distributions of the normal component of magnetic field ($|H_y|$) at the *x-z* plane for three cases of amorphous (*m*=0), half-crystallized (*m*=0.5), and crystalline states (*m*=1). It showed that the crystallization level was able to significantly change the Mie scattering resonance as well as the field confinement inside the GSST nanobars.

**Multipolar Decomposition Method.** Multipolar decomposition approach has been widely used for analyzing the optical modes being excited inside dielectric nanoantennas with various types of Mie resonances. Here, we present the detailed formulation of multipolar decomposition for analyzing their different modes as follows:[154, 173]

$$\boldsymbol{J} = -i\omega\varepsilon_0(\varepsilon - \varepsilon_d)\boldsymbol{E} \,, \tag{1}$$

where $\omega$ denotes the angular frequency of the wave, $\varepsilon_0$ denotes the permittivity of vacuum, $\varepsilon_d$ denote the dielectric constant of the surrounding medium, and $\varepsilon$ denote the dielectric constant of the particle.

Moreover, the radiation characteristics of the high index nanoparticles consist of charges and currents in terms of the internal structure. It requires the inclusion of toroidal moments, which are called the mean-square radii of the corresponding dipole distributions.[174] These physical quantities could be expressed with the Cartesian basis as the following:[154, 173]

$$\boldsymbol{p} = \int \varepsilon_0(\varepsilon - \varepsilon_d)\boldsymbol{E}d\boldsymbol{r} \,, \tag{7}$$

$$\boldsymbol{m} = \frac{-i\omega}{2}\int \varepsilon_0(\varepsilon - \varepsilon_d)[\boldsymbol{r} \times \boldsymbol{E}]d\boldsymbol{r} \,, \tag{8}$$

$$\boldsymbol{t} = \frac{-i\omega}{10}\int \varepsilon_0(\varepsilon - \varepsilon_d)[(\boldsymbol{r} \cdot \boldsymbol{E})\boldsymbol{r} - 2r^2\boldsymbol{E}]d\boldsymbol{r}. \tag{9}$$

The mean-square radii of the dipole distributions are:

$$\overline{\boldsymbol{R_m^2}} = \frac{-i\omega}{20}\int \varepsilon_0(\varepsilon - \varepsilon_d)[\boldsymbol{r} \times \boldsymbol{E}]r^2d\boldsymbol{r} \,, \tag{10}$$

$$\overline{\boldsymbol{R_t^2}} = \frac{-i\omega}{28}\int \varepsilon_0(\varepsilon - \varepsilon_d)[3r^2\boldsymbol{E} - 2(\boldsymbol{r} \cdot \boldsymbol{E})\boldsymbol{r}]r^2d\boldsymbol{r} \,, \tag{11}$$

where only the magnetic and toroidal components are considered, becasue the electric one does not contribute to radiation.[174] For the quadrupole moments, we have the following expressions:

$$\overline{\overline{\boldsymbol{Q}}}_e = \int \varepsilon_0(\varepsilon - \varepsilon_d)[\boldsymbol{r} \otimes \boldsymbol{E} + \boldsymbol{E} \otimes \boldsymbol{r}]d\boldsymbol{r} \,, \tag{12}$$

$$\overline{\overline{\boldsymbol{Q}}}_m = \frac{-i\omega}{3}\int \varepsilon_0(\varepsilon - \varepsilon_d)[\boldsymbol{r} \otimes (\boldsymbol{r} \times \boldsymbol{E}) + (\boldsymbol{r} \times \boldsymbol{E}) \otimes \boldsymbol{r}]d\boldsymbol{r} \,, \tag{13}$$

$$\overline{\overline{\boldsymbol{Q}}}_t = \frac{-i\omega}{28}\int \varepsilon_0(\varepsilon - \varepsilon_d)\big[4(\boldsymbol{r} \cdot \boldsymbol{E})\boldsymbol{r} \otimes \boldsymbol{r} - 5r^2(\boldsymbol{r} \otimes \boldsymbol{E} + \boldsymbol{E} \otimes \boldsymbol{r}) + 2r^2(\boldsymbol{r} \cdot \boldsymbol{E})\overline{\overline{\boldsymbol{I}}}\big]d\boldsymbol{r} \,, \tag{14}$$

where $\otimes$ represents the usual dyadic product. Moreover, multipolar decomposition could be implemented on the commercial simulation softwares, such as Finite-difference time-domain method (*e.g.*, Lumerical FDTD Solutions) and Finite Element Method (FEM, *e.g.*, COMSOL).[36]

Here, Figure 5b presents the multipolar decomposition simulation result of a tunable metasurface, where silicon nanostructures were placed onto the $SiO_2$ substrate with GST.[173] The unit cell design consisted of a silicon dimer so as to obtain two anti-phase electric fields. The thickness and radius of the silicon nanodisks were both fixed at 210 nm with a gap size of 50 nm. The lattice constant of the unit cell was 1 μm, while the thickness of the GST film was set to 10 nm. It showed that switching between the amorphous and crystalline states of the GST film was able to significantly change the multipolar components of the silicon dimer array antenna. Multipolar decomposition revealed the physical origin of various components, such as ED, MD, electrical quadrupole (EQ), magnetic quadrupole (MQ), *etc.*, which contributed to the total reflectance/transmittance spectra. Recently, this approach was adopted to analyze the Mie resonance characteristics of nanostructured $Sb_2S_3$ at both the amorphous and crystalline states.[71]

**Bound-States-in-The-Continuum (BIC).** The concept of BIC was originated from quantum mechanics and its localized state has an energy level being higher than the potential wells counterintuitively.[175] This BIC concept was first demonstrated experimentally in semiconductor heterostructures,[176] and it has been extended into various fields nowadays, such as acoustics,[177, 178] microwave[179, 180] and optics.[155, 181-186] For instance, BIC in optics is able to achieve sharp resonances with the designs of gratings and waveguides,[181-183] layered nanoparticles[184] and photonic crystals,[185] where the basic principle of BIC is to suppress radiative losses via destructive interferences.[187, 188] Recently, BIC has been investigated in nanophotonics with both antenna arrays[155, 186, 189-195] and individual resonators.[184, 196-198] These BIC nanoantennas are able to achieve directionality,[199] hyperspectral imaging,[194] lasing,[200-202] nonlinearity,[193, 197, 203, 204] chirality,[205] bio-sensing,[194, 195, 206] structural colors[155] and topological photonics.[199, 207]

In addition, many studies on tunable BIC resonance have been carried out recently. For example, Figure 5c presents the tunable BIC resonance enabled by the nanostructured arsenic trisulfide ($As_2S_3$), which is a photosensitive chalcogenide glass with optical properties that can be finely tuned by light absorption at the post-fabrication stage.[208] The inset of Fig. 5c presents the top-view SEM image of the $As_2S_3$ metasurface placed on a glass substrate, where the metasurface period was 490 nm and the scale bar denoted 100 nm. The resonant wavelength of this $As_2S_3$ metasurface was designed to be ~780 nm, which is below the bandgap of $As_2S_3$ (*i.e.*, 2.7 eV). The optical properties of $As_2S_3$ could be experimentally controlled via the exposure of the light with the corresponding energies above the bandgap. The refractive index of $As_2S_3$ could be tuned by the dosage of light exposure, where the maximal contrast of its refractive index due to optical annealing was ~0.08.[208] As a result, Figure 5c presents the experimental transmission spectrum for initial and exposed $As_2S_3$ metasurface. It shows that the unexposed sample supported a quasi-BIC at 780 nm and the resonance was shifted by 12 nm after the optical annealing.

**Anapole and Electromagnetically Induced Transparency (EIT).** Anapole is a dark Mie resonance with a high-quality factor, and it originates from the interference between an ED and a toroidal dipole (TD). Left panel of Figure 5d shows an actively tunable anapole resonance realized based on the nanostructured GST nanodisks.[209] To be specific, the GST nanodisks (with a radius of 1 μm and a height of 220 nm) demonstrated ED resonance at the amorphous state. Once it was switched from amorphous (a-GST) to crystalline (c-GST) via an annealing process at 145 °C, the ED resonance was then switched to anapole state due to the dramatic change in its optical contrast. In addition, EIT resonance has been demonstrated on plasmonic nanostructures,[210, 211] dielectric nanoantennas,[212] and 2D material based (*e.g.* graphene) metasurface.[213, 214] Right panel of Figure 5d schematically shows the design of a graphene metasurface, through which tunable EIT resonance was achieved by electrically controlling the effective Fermi energy levels.[213] The illumination field was the *y*-polarized terahertz waves.

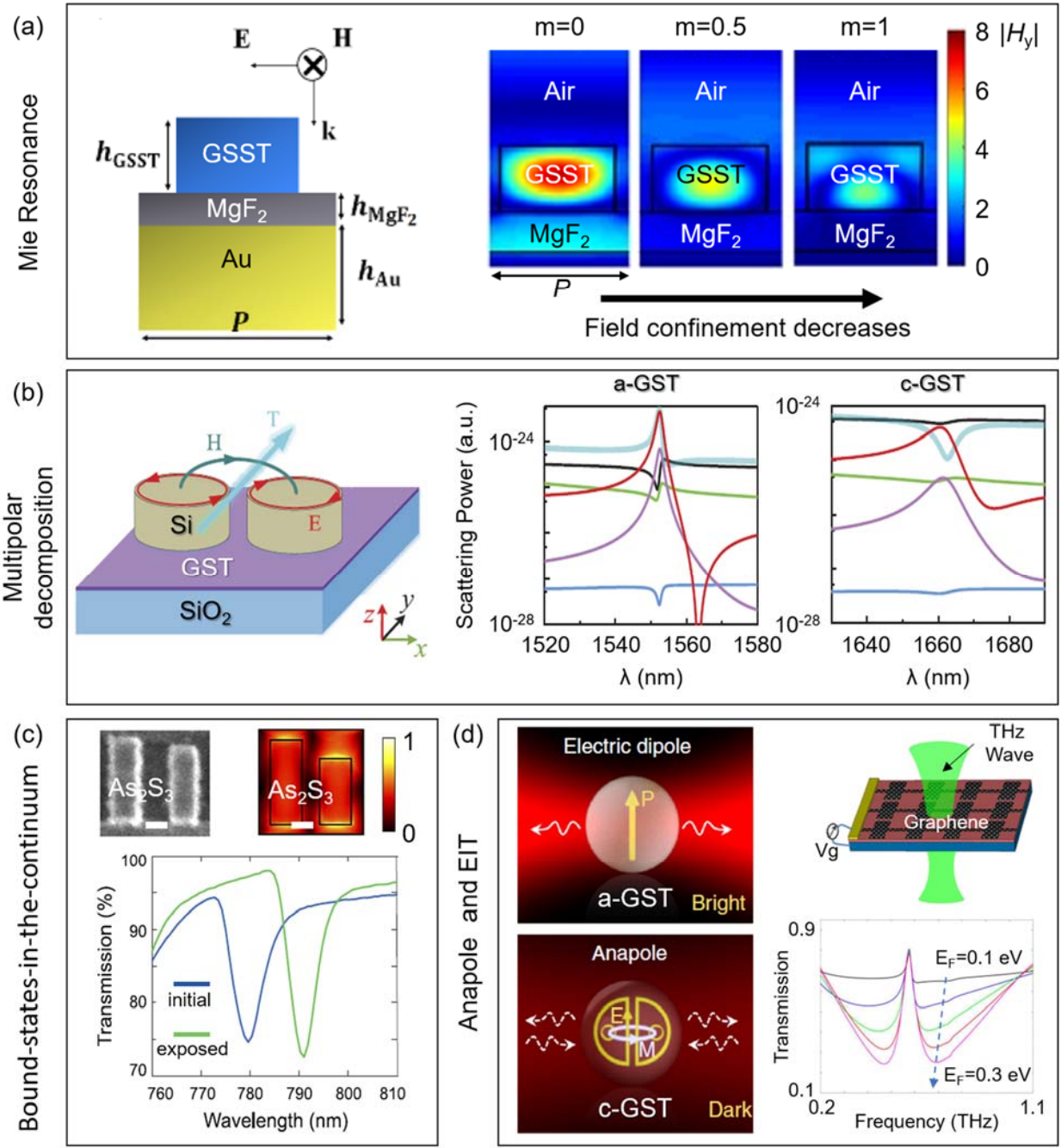

**Figure 5. Physics behind tunable metasurfaces.** (a) Tunable Mie resonance by GSST nanobar arrays with different crystallization levels at the resonance wavelength of 1.55 μm. Left panel: schematic of a unit-cell of GSST nanobar placed on a stack of MgF$_2$-Au backmirror. h$_{GSST}$, h$_{MgF2}$ and h$_{Au}$ denote the thickness of GSST, MgF$_2$ and Au film layers, respectively. $P$ denotes the pitch of the unit-cell. Right Panel: Near-field distributions of the normalized component of magnetic field(|H$_y$|) at x-z plane for amorphous ($m$=0), half-crystallized ($m$=0.5), and crystalline states ($m$=1). Reprinted with permission from the reference.[172] Copyright 2018 the Optical Society of America (OSA). (b) Multipolar decomposition method. Left panel: schematic of a hybrid PCM-dielectric metasurface consisting of silicon nanostructures placed atop of a layer of GST. Right panel: Total scattered power and contributions of different multipoles at a-GST ($m$=0)

and c-GST ($m$=1), where $m$ denotes the crystallization ratio of GST film and ranges from 0 to 1. It showed that switching the crystalline ratio of the GST film was able to change the multipolar components of the silicon dimer array antenna significantly. Reprinted with permission from the reference.[173] Copyright 2020 the Optical Society of America (OSA). (c) BIC resonance. Top panel: top-view SEM image of the $As_2S_3$ metasurface placed on a slide glass substrate. The height of $As_2S_3$ nanostructures is 180 nm and the metasurface period is 490 nm. The scale bar denotes 100 nm. Bottom panel: experimental transmission spectrum for initial and exposed $As_2S_3$ metasur-face. Reprinted with permission from the reference.[208] Copyright 2019 the Optical Society of America (OSA). (d) Tunable anapole by the nanostructured phase-change alloy GST and tunable EIT resonance by graphene metasurface. Left panel (top and bottom): shifting between scattering bright and dark states in a GST nanodisk. It shows that the resonant mode of the nanostructured GST nanodisks with a radius of 1 μm and a height of 220 nm could be switched from the electrical dipole mode to the anapole mode. Reprinted with permission from the reference.[209] Copyright 2019 Springer Nature. Right panel: schematic of structural design of the graphene metasurface (top) and the measured transmission (bottom). It has been demonstrated to achieve the tunable EIT resonance by tuning the effective Fermi energy based on the electrical voltage, where the transmission spectra could be actively controlled. Reprinted with permission from the reference.[213] Copyright 2019 Elsevier.

## Design Methodologies.

Although tunable metasurfaces progressively become versatile and powerful, the design methodologies are currently suffering from two bottlenecks. For instance, the existing phase engineering design approach cannot sustain the fast-growing demand for applications operating in broadband wavelength. In addition, conventional design methods typically include model designs, trial-and-error method, parameter sweep, and optimization algorithms. Performing full wave numerical simulation assisted by parameter sweep and optimization algorithm usually requires plenty of computational resources and time. If the design requirement changes, simulation must be repeated and validated manually till it meets the goal. Hence, there is an increasing demand to develop fast, efficient, accurate and automated design approaches to address the needs for tunable metasurface design.

**Inverse design** aims to achieve the optimum custom-defined performance. To accomplish this task, different inverse design optimization techniques have been developed, including particles swarm optimization (PSO) and genetic algorithm (GA). Figure 6 summarizes typical ML-assisted design methodologies, which have been adopted for tunable metasurface design. GA is a type of gradient-free algorithm that originated from the idea of biology. In each iteration step, GA generates many solutions called population, inspired by biological operators, such as mutation, crossover and selection. Solutions obtained from each iteration will be scored based on objective function or fitness function. Stop criteria could be converged to the optimum solution by defining a certain number of generations. Wallace *et al.* adopted the GA approach to design an efficient grating tunable laser.[215] Figure 6a shows the GA flow chart of the inverse design optimization. They demonstrated a nearly doubling slope efficiency of the laser using the mutation probability of 0.2 and the crossover probability of 0.15. Thureja *et. al.* adopted iterative GA provided by MATLAB toolbox and devised an ITO metasurface for efficient beam steering.[3] The directivity was enhanced by ~84% compared with the forward design approach. The input data to the GA model was the tunable phase-amplitude at each applied voltage with the crossover probability of 0.95 and optimization run of 250 generations.

PSO is a typical swarm-intelligence algorithm that imitates the collective behavior observed in social animals such as birds and fish.[216] The decentralized system can be self-organized to find the globally optimal solution to a problem, with a proper interplay of exploration and exploitation. PSO consists of a group of particles in a swarm and the position of each particle is an *N*-dimensional vector, which represents all parameters of the design space to be optimized. because those particles move together around the design space, their positions will converge after multiple iterations to reach the optimal decision. Chung *et. al.* utilized the PSO technique to boost the beam steering with the angular deflection from < 30° to 132° in an electrically tunable metasurface.[217] PSO has the advantage of performing adjoint-based local-optimization design iterations inside a global-optimization search. The adjoint function was based on the switching efficiency and the radiated electric fields of the tunable metasurface with and without applying voltage. The proposed tunable metasurface consisted of two layers of $TiO_2$ grating on the front and back sides of the device and one silicon grating embedded inside the LCs (Fig. 6b). The global optimization parameters consisted of the thickness of each layer and the intermediate layers of ITO and alignment marks as well. After the 100th iteration of optimization, the switching efficiency reached 60% which was two times higher than that achieved using conventional design method.

Novel models of GA have been developed to tackle large-scale optimization problems. Allen *et. al.* utilized the multi-objective GA optimization technique to efficiently optimize four-layer metasurfaces operated in GHz range.[218] The fitness function was based on the physical limit of in-band reflection, where the theoretical reflection was limited by the Bode-Fano bound. In addition, adaptive GA was proposed by Jafar-Zanjani *et. al.* to tackle four complex design problems for optical metasurfaces and each case suffered from computational and fabrication challenges using conventional design methods.[219] For example, a binary pattern plasmonic array exhibited high tolerance on fabrication imperfections and high reflection efficiency for beam-steering purposes; a compact birefringent all dielectric metasurface with fine pixel resolution outperformed canonical nano-antennas[220], *etc*. Eventually, they showcased the advantages of the combination of binary-pattern metasurfaces with the adaptive GA technique. Jin *et. al.* established a segmented hierarchical evolutionary algorithm, a subset of GA, to achieve a rapid converging speed over a small number of iterations.[221] Owing to the improved fitness function based on coherent coefficients, they successfully solved large-pixelated, complex inverse meta-optics design and fully demonstrated the targeted performance of broadband, high image fidelity and full color holographic metasurfaces.

The need for a fast solution to solve a complex problem with a giant number of design parameters led to the birth of AI in 1956.[222] AI is a technique that enables machines to do classifications and decisions efficiently without human interactions. In order to make a decision properly, AI requires a large amount of input data as a training data set to learn its model efficiently. ML came into operation in the late 80s and it is a subset of AI that uses statistical methods to train the model efficiently with more input data. ML requires manually assigning the most important features of the problem to do the classification efficiently. It can interpret the decision behind ML easily, based on the importance of parameters in the decision tree. Li *et. al.* incorporated an ML technique to achieve a real-time reprogrammable digital metasurface imager operated in GHz range.[5] In this work, the training model was based on the principal component analysis (PCA). Figure 6d illustrates the flow chart of the training model of the reprogrammable imager, where the recorded images were used as training samples to refine PCA parameters for accurate prediction. Then, the reconstructed images were stored as a training data set, where the whole training process time was 20 min. Next, the trained ML model was tested to monitor another moving person. They showed that the ML imager trained by PCA was able to produce a high-quality image with only 400 measurements, far less than the number of 8000 unknown pixels.[215]

ML-assisted design can be divided into two categories: inverse design and forward design. They are expected to expedite the discovery and design of novel metasurfaces with custom-defined performance. For inverse design, the EM response and spectrum are defined as an input, and the corresponding structure (size, shape, *etc*.) can be predicted quickly. In contrast, the geometry parameters of meta-atoms are defined as input in the forward design, and the EM response of metasurfaces can be predicted without running EM simulation.

Neural Network (NN) is a subset of ML that is inspired by the functionality of human brain cells, which are called neurons. NN does not need to assign any feature for doing classifications. Instead, it is based on self-learning at multiple levels of abstractions. NNs can automatically find the most important features to do classification, followed by efficiently performing decisions. NN performance increases with the amount of the input data set. An *et. al.* introduced a deep neural network (DNN) approach to significantly improve both speed and accuracy for optical metasurfaces characterization.[4] They showed a good performance in predicting the phase of meta-atom and an efficient design of reconfigurable metasurfaces. Figure 6c shows the DNN model to train and inversely design the accurate phase of meta-atom. The input parameters to the DNNs model were structure geometries, period, and permittivity versus wavelength, while the output parameters included the amplitude and phase of the transmitted light. After being trained with 14,800 groups of meta-atom data sets, the predicting DNN was able to achieve >99% accuracy for both amplitude and phase prediction. These results show the potential of DNN in achieving fast and accurate on-demand metasurface design.

Lenaerts *et. al.* developed a NN model to optimize transmission and reflection from Fabry-Perot resonator and Bragg reflectors.[223] The input parameters included the Finesse coefficient and the phase difference between different transmitted partial waves. They significantly improved the NN model's efficiency by adding the Fourier transform of the desired spectrum to the optimization procedure. Taghvaee *et. al.* has achieved fast and accurate prediction of radiation pattern for 6G networks using NN.[224] Owing to the convolutional NN (CNN) model, they achieved an accuracy of up to 99.8% compared with full wave simulation. After training, their NN model was able to accurately predict a radiation pattern faster than full wave simulation program. Thompson *et. al.* has incorporated an artificial NN model, which was called sum product network (SPN), to efficiently select the best switchable mirror design with a mean average error of less than 2%.[225] SPN enabled a model faster, and more accurate than the standard DNN. Moreover, they extended their model from 1D to 2D GST grating, which exhibited reflection and transmission efficiency of 75%~95%.

**Forward Design.** Conventional trial-and-error approach using EM simulators to achieve the required performance is time consuming. The development of theoretical accurate models can offer a fast and efficient solution, but the theoretical models are hard to obtain for most of metasurface applications. Recently, Sperrhake *et. al.* has developed a semi-analytical scattering matrix model to estimate the response of Fabry-Perot cavity with high accuracy and it is faster than EM simulators (Fig. 6e).[226] It started with a scattering matrix that contained Jones-matrices of forward and backward light propagation coefficients. Then, it adopted a semi-analytical stacking algorithm (SASA) to accurately estimate the transmission and reflection of any arbitrary number of stacked layers. The results matched well with experimental data, which could expedite the design process of stacked metasurface layers. On the other hand, a circuit model has been applied in the GHz range to explain the performance of transmission lines, waveguides, *etc*. Rastgordani *et. al.* has developed it to predict the performance of a broadband absorber in visible and THz ranges.[227] This model offered accurate calculations faster than EM simulators and the results agreed well with the experimental data.

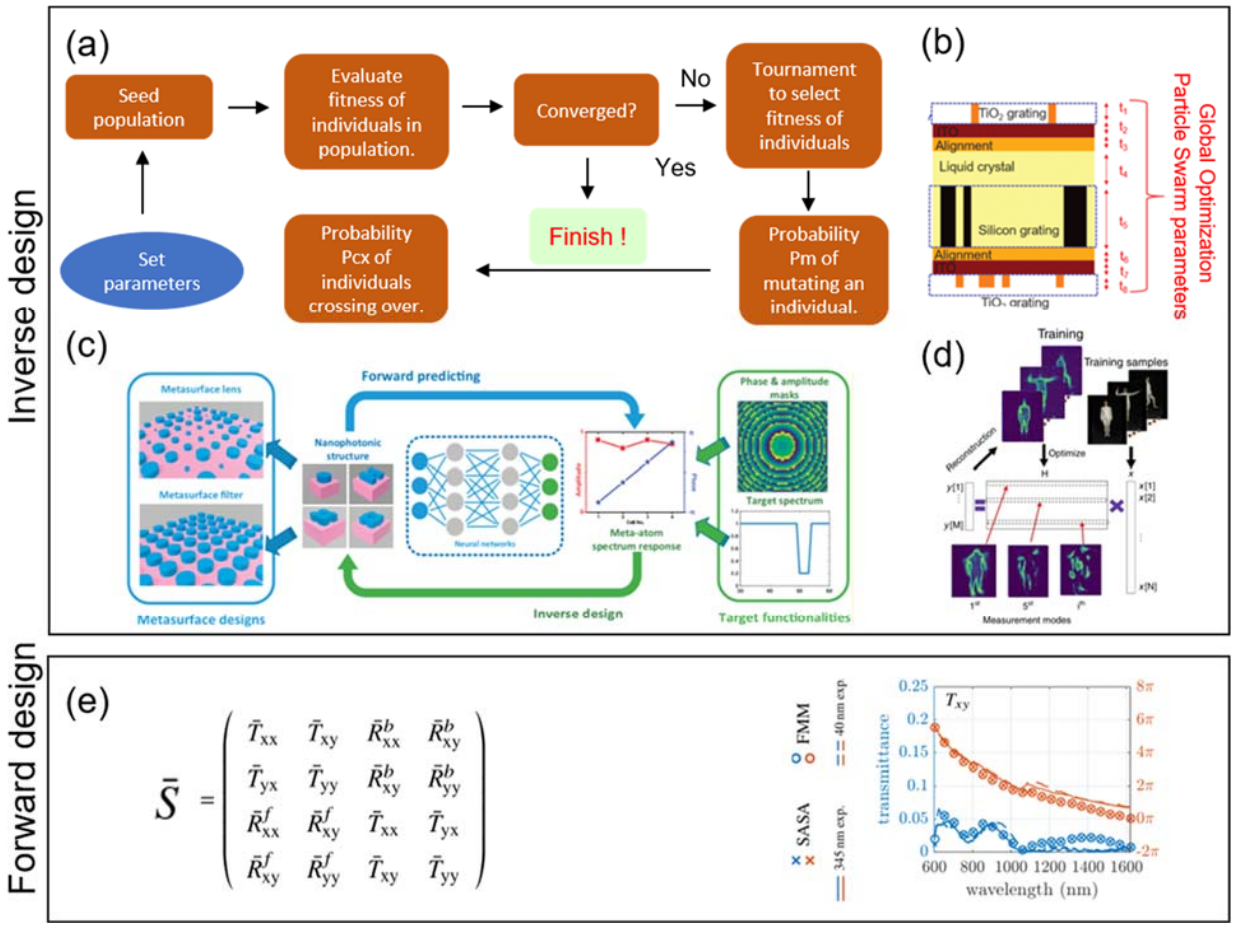

**Figure 6. Machine learning-assisted design methodologies**. (a) Flow chart of optimization flow based on genetic algorithm for designing NIR lasing in p-type InP grating array.[215] Reprinted with permission from Optica Publishing Group, Copyright 2020. (b) Schematic diagram of tunable metasurface based on LCs for beam steering application. Each layer thickness selected and optimized based on the particle swarm optimization method.[217] Reprinted with permission from American Chemical Society, Copyright 2020. (c)

Design flow of dielectric metasurface based on a deep learning approach, which consisted of neural network with four hidden layers with four input design parameters and 31 output parameters for desired output metasurface spectrum for metalens and metasurface filter applications.[4] Reprinted with permission from American Chemical Society, Copyright 2019. (d) Training flow chart of machine learning-based reprogrammable metasurface for imaging applications using a principal component analysis approach.[5] Reprinted with permission from Springer Nature, Copyright 2019. (e) Left panels shows analytical formula of scattering matrix based on Jones matrix for estimating transmission and reflection of stack of gold nanowire metasurface embedded in glass. Right panel shows good matching between experimentally measured transmission and analytical formula.[226] Reprinted with permission from Optica Publishing Group, Copyright 2019.

## APPLICATIONS

Tunable metasurfaces based on resonant subwavelength photonic structures with active constituent materials, hold a huge promise for the realization of full control of EM fields in terms of phase, amplitude, and polarization at the nanoscale. The development of tunable metasurfaces has been driven rapidly to embrace the increasing demand on optics and nanophotonics towards real applications, and the ultimate goal is to realize ultrafast dynamic control at the individual pixel level. In this section, we will summarize recent advances in major applications such as dynamic beam focusing, wavefront shaping, beam steering and image display, which are illustrated from Figures 7-9.

**Dynamic Beam Focusing.** The flatness of metalens makes it feasible to manufacture compact optical systems. Although the number of publications on beam focusing on the broadband spectrum has rapidly increased during the past decade, its progress towards commercial devices such as a tunable zoom lens in the camera, is still lagging behind anticipation. The challenges typically come from the limited dynamic control mechanism and the slow response speed, which hinder its integration into an imaging system to replace the traditional zoom lens array.

Various external stimuli have been explored to induce the tunability of metasurface, which are largely dependent on material properties and tuning mechanisms. Mechanical stimulus was used to continuously change the focusing length of a metalens in visible frequency by stretching the PDMS substrate. A 1.7× zoom metalens was realized with a varifocal length over a range of 100 um, as shown in Fig. 7a.[28] However, it applied the mechanical force to the whole metasurface, which was not able to individually control each pixel using this approach.

To improve the performance, an electrically tunable adaptive metalens was proposed to continuously control the focal length, astigmatism and shift. It was realized through engineering the strain field of substrate by applying an electrical bias to dielectric elastomer actuators (referred to as artificial muscles), as shown in Fig. 7b.[228] It opens the door to future miniaturization of an imaging system for cell phone and wearable displays. Figure 7c schematically shows the concept of varifocal silicon metalens, which relied on electrically controlling the refractive index environment of silicon metalens by means of a resistor embedded into a thermo-optical polymer.[229] This work exhibited a continuously varying focal length larger than the Raleigh length with a low driving voltage in the visible frequency. In another work, achromatic varifocal metalens consisting of TiO$_2$ nanopillars was designed and fabricated, as depicted in Fig. 7d.[230] The wavelength- and polarization-dependent metalens was able to change its focal length from 220 μm to 550 μm over a wide visible spectrum from 483 nm to 620 nm upon the incident polarization rotation.

In the MID-IR region, PCMs provide a promising approach for reconfigurable metasurfaces. The extremely large refractive index contrast between amorphous and crystalline states and the relatively low loss in the mid-IR region empower PCMs metasurfaces high efficiency, high Q factor and multifunctionalities. As shown in Fig. 7e, a high performance varifocal metalens based on GSST was demonstrated to operate at the wavelength of 5.2 μm with diffraction-limited performance.[231] The switching contrast of this bi-state metalens reached 29.5 dB and a mid-infrared optical image comparable to traditional volume zoom optics was realized, which will be beneficial to mid-infrared optical imaging, sensing and display applications.

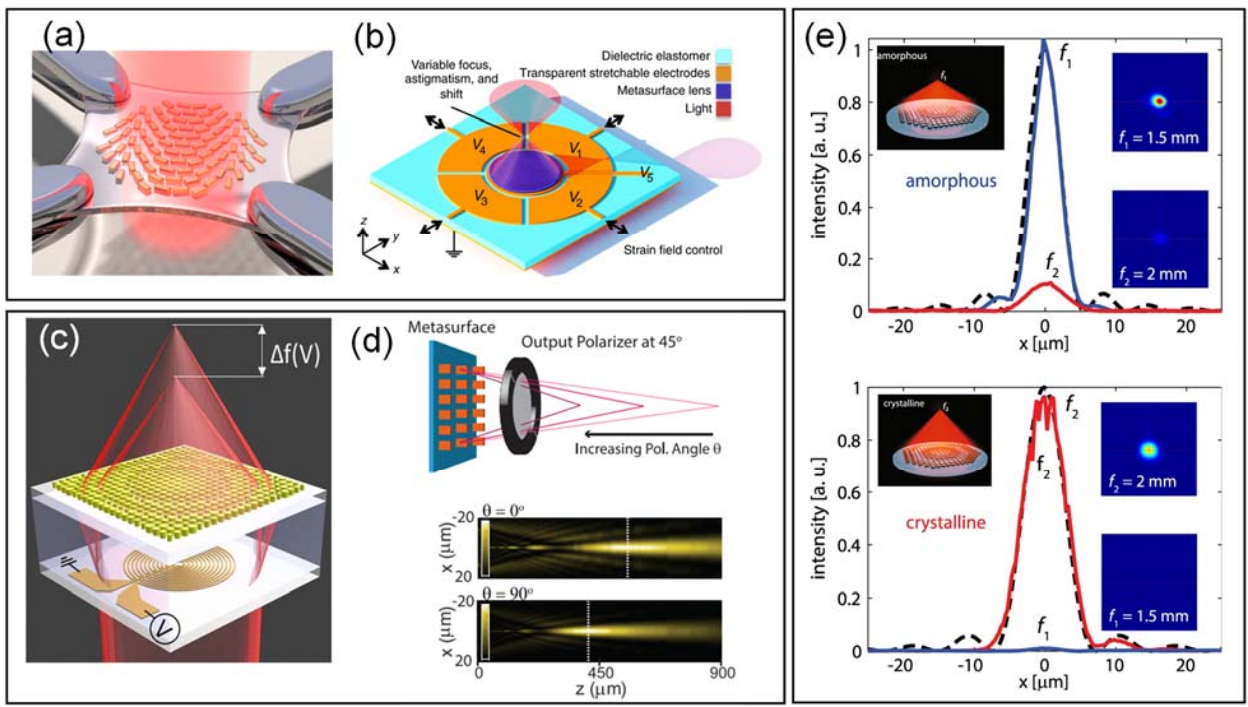

**Figure 7. Dynamic beam focusing**. (a) Schematic illustrations of tunable metalens on a stretchable PDMS substrate.[28] Reprint with permission from American Chemical Society, Copyright 2016. (b) A schematic of an adaptive metalens in which a dielectric elastomer actuator (DEA) with five addressable electrodes were combined to allow for electrical control over the strain field of the metasurface.[228] Reprinted with permission from AAAS, Copyright 2018. (c) Schematic of the concept of electrically controlled varifocal metalens, which relies on dynamically controlling its refractive index environment by means of an electric resistor embedded into a thermo-optical polymer.[229] Reprinted with permission from American Chemical Society, Copyright 2018. (d) Schematic of the operation principle of operation of an achromatic varifocal metalens. Linearly polarized light passed through a metalens and an output polarizer oriented at 45°. The focal length changed as the input polarization was rotated.[230] Reprinted with permission from American Chemical Society: Copyright 2019. (e) Binary switching of metasurface with GSST, Top panel: schematics of a reconfigurable varifocal metalens with a dimension of $1.5 \times 1.5$ mm$^2$. Incident light was focused on the first focal plane ($f_1 = 1.5$ mm) when the meta-atoms are in the amorphous state (left) and the second focal plane ($f_2 = 2.0$ mm) in the crystalline state (right). Bottom panel: optical characterization of focal spot profiles for the metalens in two states: amorphous (left) and crystalline (right).[231] Reprinted with permission from Springer Nature, Copyright 2021.

## Wavefront Shaping and Beam Steering.
Dynamic control of the beam wavefront is one of the grand challenges faced by traditional optical systems for various applications, such as adaptive optics, LiDAR, imaging display, communication, *etc*.[27] Tunable metasurfaces have been viewed as an extraordinary promising solution to achieve full control of the characteristics of light beam.

Photonic spin-orbit interactions have recently attracted much attention due to their similarity with the spin-orbit interaction of electrons in solids.[37] Dynamic PCM metasurfaces offer a rich degree of freedom to manipulate the spin and orbital angular momenta of light. Annealing GST at different temperatures with proper timing, it was able to gradually change its refractive index among different phase states (amorphous, intermediate, and crystalline states). Zhang *et al*. utilized the Pancharatanam-Berry (PB) phase and propagation phase of GST nanofins to realize the conversion from spin angular momentum (SAM) to orbital angular momentum (OAM).[232] The SEM image of the fabricated GST nanofins is shown in the top panel of Fig. 8a. Note that the PB phase remained unchanged across different phase states, while the propagation phase varied accordingly. The bottom panel of Fig. 8a displays the measured diffraction patterns at different crystallization levels under left and right circular polarized (LCP and RCP) light together with interference patterns obtained through the interference of the OAM beams under a titled circularly polarized Gaussian beam.

Recently, an all-dielectric active metasurface based on III-V MQW nanostructure was demonstrated to support a hybrid Mie-guided mode resonance (Fig. 8b), which achieved relative reflectance modulation of 270% and a phase shift from 0° to 70°.[233] The beam steering was realized by applying an electrically bias on each subwavelength element. In addition, an electrically reconfigurable non-volatile GST metasurface achieved spectral tuning of light scattering in the visible and NIR regime.[69, 73] Its device configuration and switching schemes are shown in Fig. 8c.[69] The giant refractive index change of GSST has great potential for on-demand phase and amplitude control at individual pixels for active metasurface. Furthermore, a high-efficiency silicon-nanodisk metasurface infiltrated with LC was demonstrated, as shown in Fig. 8d. It realized the switching of a laser beam from the angle of 0° to 12° together with 50% tuning efficiency.[234] Beam steering at transmission mode will open up new opportunities for full range scanning of LiDAR applications.

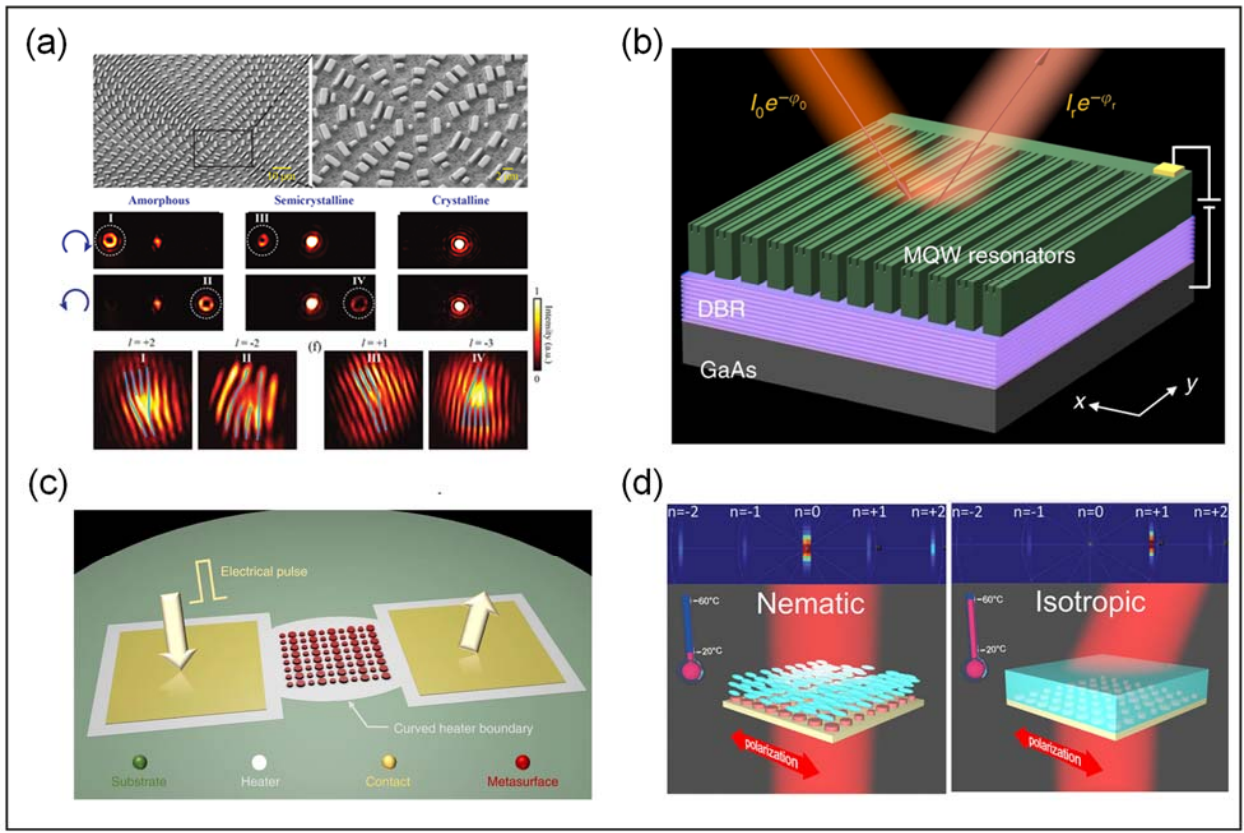

**Figure 8. Wavefront Shaping and Beam Steering.** (a) Up panel: the SEM image of the fabricated GST nanofins on top of gold. Bottom panel: Measured $\pm 1^{st}$ and $0^{th}$ diffraction modes of the SAM-to-OAM converter with different phase states (amorphous, semicrystalline, crystalline) under left- or right-circular polarized light (top and middle). In amorphous states, the converter can generate OAM with $l = \pm 2$ charges, while in the semicrystalline state, it displays $l = +1$ and -3 OAM [232]. Reprinted with permission from Wiley-VCH GmbH, Copyright 2020. (b) Schematic of all-dielectric MQW metasurface for beam steering. The metasurface consisted of an $n$-doped GaAs substrate, a distributed Bragg reflector (DBR), and a 1230-nm-thick MQW layer. There was a 50-nm-thick $p$-doped GaAs layer with a doping level of 1019 cm$^{-3}$ that was grown on top of the MQWs as a top contact.[233] Reprinted with permission from Springer Nature, Copyright 2019. (c) Schematic of the electrothermal metasurface switching configuration using GSST for reflection mode beam steering.[69] Reprinted with permission from Springer Nature, Copyright 2021. (d) Dynamic beam switching by all dielectric metasurfaces composed of Si nanodisks infiltrated with LCs. Concept of beam switching metasurfaces infiltrated with LCs at nematic state; Left panel: nematic state. Right panel: isotropic state.[234] Reprinted with permission from American Chemical Society, Copyright 2018

**Image Display**. Holographic metasurface uses a concept similar to traditional optical holography to recreate a complex optical wavefront. However, the resolution of metasurface holographic images is not comparable to traditional digital holograms due to the limited bandwidth of metasurface hologram. OAM holography recently achieved multiplexing of up to 200 independent channels by complete and independent amplitude and phase manipulation in a complex metasurface.[235] Figure 9a shows the working principle of ultrahigh-dimensional OAM-multiplexing holography. The image information loaded in different orbital momentum modes was extracted by a Fourier transform to allow holographic video display for new applications in smart and compact wearable image display.

The hologram is natively suitable for 3D images visible to the naked eyes. It is a long-standing challenge to achieve dynamic holography metasurface in the visible range with a large frame rate. Figure 9b shows a design of meta-holography, consisting of silicon nitride nanopillars with a high-speed dynamic structured laser beam modulation module, which achieved $2^{28}$ different holographic frames at a high frame rate of 9523 frames per second in the visible range.[236] This 3D holographic display method can satisfy the requirement of potential applications in laser fabrication, optical data storage, optical communications and information processing.

Various metasurface designs have demonstrated multi-channel holographic images display.[237, 238] However, these are still passive technologies, and the optical characterization are fixed during the manufacturing process. Two conceivable approaches for active metasurface hologram display based on LC platform showed the real time holographic images.[239, 240] Figure 9c shows the schematic illustrations for the LC-integrated metasurfaces.[239] The LC modulator provides versatile external stimuli to realize holographic images, for example, electric field, heat and surface pressure. Such a dynamic meta-holographic platform will provide a path for compact display devices in the application of tunable lens, augmented reality/ virtual reality (VR/AR) display and LiDAR applications.

Metasurface as a promising nanophotonics concept has been demonstrated in various imaging systems for the replacement of conventional bulky optical components, ranging from chiral imaging, optical coherence tomography, fluorescent imaging, super-resolution imaging, quantitative phase imaging, *etc*.[241] One recent work demonstrated a spin-multiplexed metasurface based optical imaging system.[242] Figure 9d shows the schematic of the concept for spin-dependent function control. The all dielectric metasurfaces performed a 2D spatial differentiation operation and thus achieved isotropic edge detection. In addition, the metasurface was able to provide two spin-dependent, uncorrelated phase profiles across the entire visible spectrum. Therefore, based on the spin-state of the incident light, the system could be used for either diffraction-limited bright-field imaging or isotropic edge-enhanced phase contrast imaging. The ultrathin architecture of the planar metasurface provides a platform for the scalability and integration of biomedical microscopy imaging systems.

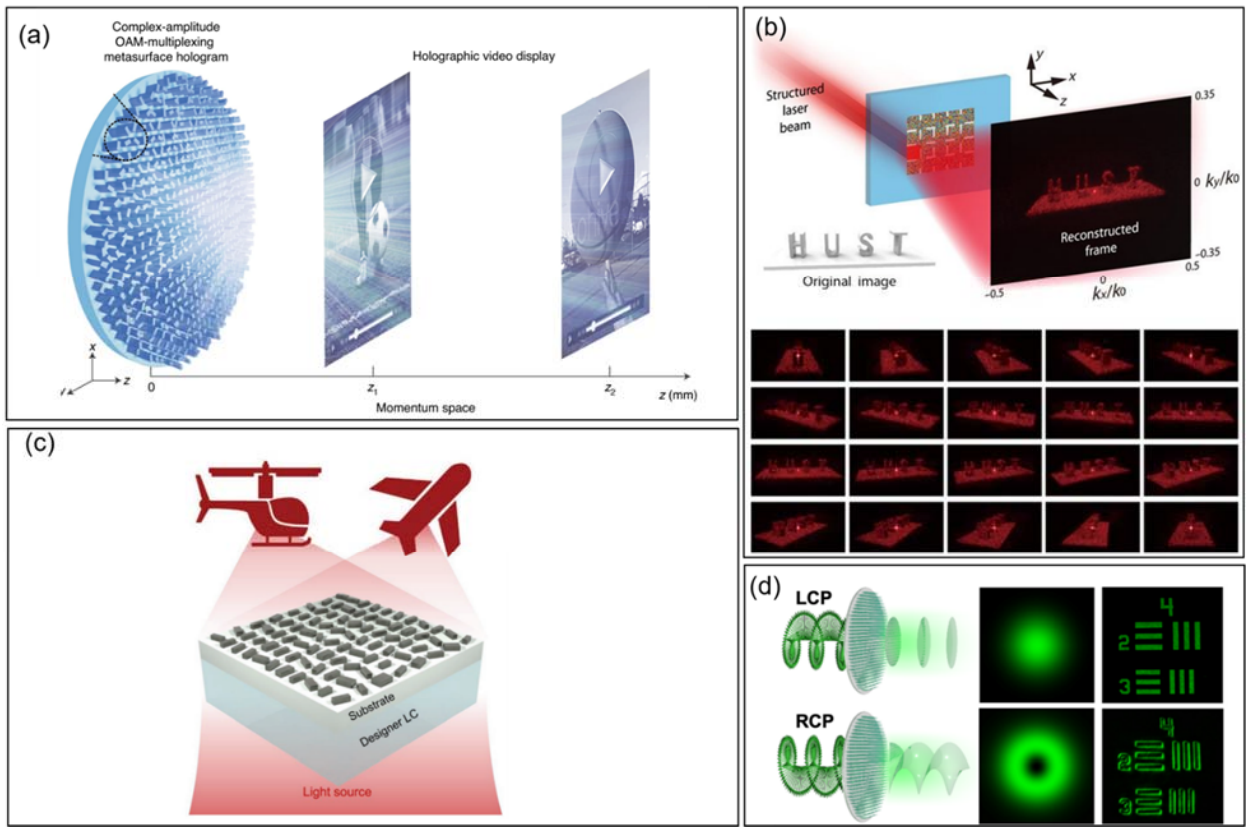

**Figure 9. Image Display.** (a) Design principle of ultrahigh-dimensional OAM-multiplexing holography based on a large-scale complex-amplitude metasurface. Time-dependent image frames encoded in an OAM signature with continuous helical mode indices ranging from -50 to +50. Metasurface offers independent control of both amplitude and phase responses of transmitted light. Optically addressable holographic video display at two image planes ($z_1$ and $z_2$) by addressing a large number of OAM-dependent orthogonal image frames.[235] Reprinted with permission from Springer Nature, Copyright 2020. (b) Design and experimental

results of high-speed dynamic space channel multiplexing meta-hologram based on silicon nitride nanopillars. Structured laser beam opened a specific space channel in the designed sequence, and continuous frames of a holographic video are displayed.[236] Reprinted with permission from AAAS, Copyright 2020. (c) Schematic illustrations for the LC-integrated metasurfaces for ultracompact dynamic holographic displays. Specifically designed LCs and efficient helicity-encoded meta-holograms are combined to realize stimuli-responsive dynamic displays.[239] Reprinted with permission from Wiley-VCH GmbH, Copyright 2020 (g) Schematic of the concept for spin-dependent function control metasurface in the application of biomedical microscopy imaging.[242] For light incident on the device with LCP, the metasurface imprinted a masking function ($M_1$) on the output beam resulting in a constant phase profile and a Gaussian intensity distribution and flipped the handedness of the incident polarization. For light incident on the same device with RCP, the metasurface imprinted another masking function ($M_2$), resulting in a spiral phase profile and a donut-shaped intensity distribution, and again flipped the handedness of the polarization. Reprinted with permission from American Chemical Society, Copyright 2020.

## CONCLUSIONS AND OUTLOOK

In conclusion, we have briefly illustrated the progress and exciting achievements on tunable/reconfigurable metasurfaces based on various approaches controlling the interaction of EM and meta-atoms. In this review, we focus on active materials and their tuning mechanisms, theory and new physics, and major applications. Because of their superior properties to conventional optics, this field evolved rapidly over the last decade which has been largely driven by the increasing demand for next generation nanophotonics for various applications including emerging areas such as AR/VR, LiDAR, smartphone, autonomous vehicles, *etc*. Traditional tunable components such as SLM using LCs and digital micromirror device (DMD) based on MEMS technology exhibit pixel resolution of several microns and above, which is much larger than the optical frequency. In contrast, thanks to advanced nanofabrication technologies, tunable metasurfaces offer a revolutionary platform towards the full control of polarization, amplitude and phase of EM waves with subwavelength pixel size. This key advantage is vital for applications, such as display and LiDAR, because tunable metasurfaces based SLM with subwavelength pixel resolution are capable of getting rid of undesirable diffraction orders. On the other hand, even though we have witnessed remarkable accomplishments in this field, there are still some trade-offs to be overcome. For example, the induced tunability relies on active materials and novel tuning mechanisms, which inevitably require complex configuration and external control units to be integrated into the device and system. Furthermore, active materials offer more degrees of freedom for complex functionalities, but their absorption may deteriorate the efficiency of active devices, making them uncompetitive compared with passive counterparts. Therefore, much more effort is still required to push this technology to the next level with high performance as expected, including faster switching speed, larger tuning range and modulation depth of phase and amplitude, higher efficiency/lower loss, lower power consumption, *etc*. In this context, we analyze the potentials of those active materials and tuning mechanisms. Based on that, we provide perspectives on the development and opportunities in near future.

**Active materials** that constitute tunable metasurfaces are the fundamental building block to acquire tunable functionality. Among them, non-volatile chalcogenide PCMs stand out due to their unique properties, such as drastic optical contrasts, fast switching speed and long-term stability. So far, the most critical effort towards of reversible switching/fully reconfigurable of PCM metasurfaces is to develop a reliable and repeatable approach for the phase conversion between their amorphous and crystalline states. The phase conversion typically can be made by either bulk thermal heaters or focused beam from a pulsed laser with a relatively large beam size covering the entire surface of nanostructures. Note that most previous works only demonstrated one-way switching, meaning that only full or partial crystalline of initial amorphous PCMs was completed. Recently, several groups have demonstrated optically and electrically reversible switching of PCMs, *i.e.*, two-way switching between amorphous and crystalline states. For example, rewritable sub-diffraction resolution color printing over a thin film of $Sb_2S_3$ was demonstrated with femtosecond laser pulses,[70] electrical pulses were applied to trigger and complete the two-way phase conversion of GST (nanobeam)[73] and GSST (nanostructures).[69] Currently, both optical and electrical switching become accessible for phase conversion of PCMs; however, a lot of effort is still required to explore their feasibility for individually addressing subwavelength PCM nanostructures to achieve pixel-level control of EM response. In fact, this is a grand challenge not only applicable for PCM metasurfaces, but also the entire field of metasurfaces. To realize this goal, it requires applying short electric pulses or subwavelength resolution laser pulses as external stimuli onto individual unit cells of metasurfaces. This is a challenging task and upon its accomplishment, it would be a milestone in the development of nanophotonics and bring us to an era of on-demand photonics.

With the rapid development of neuromorphic computing, neuromorphic photonics has emerged as a promising hardware platform to address the needs of ultrafast switching, low power, large bandwidth and better interconnectivity. At present, tunable metasurfaces and photonics integrated circuits (PICs) are the two well-established nanophotonics platforms for free space and on-chip light manipulation, respectively. Merging these two platforms has been observed to make a paradigm shift to develop large scale all optical NN on a chip, which outperforms the electronic approaches with higher speed, broader bandwidth and lower power.[243]

Semiconductors and TCOs based metasurfaces are very promising due to their relatively high readiness for CMOS compatible processes allowing for mass production and the well-known optical and electronic properties. Concerns for some materials arise from their dielectric breakdown strength that limits the amount of electrical bias and their bandgaps that limit optical performance in the visible region. Besides those well-established semiconductors having been exploited for tuning, MQW semiconductor heterostructures have also been studied due to their electrically modulated energy and strength of intersubband transition. Previous work showed that its minimum refractive index modulation was ~10 ns, which was several orders of magnitude greater than that of the free-carrier effect.[244] A more recent work demonstrated the modulation of a hybrid Mie-guided mode resonance via a quantum-confined Stark effect, which is a novel tuning mechanism.[245]

LCs are quite mature because of the multiple modulation stimuli and well-established fabrication methods, making them low-hanging fruit for tunable metasurfaces. However, due to their intrinsic material property, it is not achievable to individually address each LC cell at the nanoscale; the thickness of LC cells is a hindrance for device miniaturization and high performance; its response time/switching speed is restricted at milliseconds level. These serious issues have limited the deployment of LCs based metasurfaces in active functional devices and their practical applications. A huge effort is needed to make a breakthrough for this technology.

Ultrathin materials represent a large family of various materials, which provides a new platform to explore novel tuning mechanisms for metasurfaces. When the thickness of a layer of TCO was reduced to the range from several nm to a few tens of nm, the quantum size effect became significant to enable reversible shifting of SPR.[246] Thickness-dependent optical properties have also been observed and studied on a thin flake of layered Ruddlesden–Popper perovskites (RPPs). They exhibit the optimal thickness in the range of 20 nm – 60 nm for non-linear optical effects and third-harmonic generation (THG).[247] In terms of 2D materials, besides graphene, there exists a great variety of monolayer TMDCs, which have also manifested exotic and tunable optical properties. By twisting or stacking two or more monolayers of TMDCs via VDW force, the bandgap of the interlayer can be tuned, breeding an emerging class of materials, *i.e.*, twisted VDW materials. Recently this idea has been experimentally manifested by a magic-angle laser in such materials.[248] Twisted bilayers of photonic graphene lattice patterned on InGaAsP MQW gain material resulted in nanocavities with strong field confinement and a high-quality factor. It revealed that the confinement mechanism of magic-angle lasers does not rely on a full bandgap but the mode coupling between the twisted bilayers. This opens up a new revenue to tailor photonic response with artificially engineered VDW heterostructures. This emerging area will extend the concept of "photonic materials with a twist" to tunable metasurface design, which may boost enormous applications in future.[249]

**Tuning mechanisms** tie up tightly with the modulation of optical properties of active materials upon various stimuli and each of them exhibits merits and demerits. PCM based metasurfaces can be switched in either direct thermal, electro-thermal or opto-thermal manner. However, it requires a constant supply of heat, or electrical/light pulses as the phase transition temperature of most of PCMs are quite high. Their phase change process must be adaptive to ambient temperature, which is a pre-requisition to find suitable applications. MEMS-based tuning approach has been recognized that the switching speed is limited by its mechanical components, which is generally slower than other types of tuning approaches. Meanwhile, its durability and lifetime are affected by the mechanical moving parts. In comparison, optical pumping-based approaches hold great promise for ultrafast switching of EM responses of metasurfaces. Their tuning mechanisms rely on various light-induced effects that include free-carrier generation, opto-thermal effect and nonlinearities. Due to the intrinsically fast carrier dynamics involved in these photo-effects, ultrafast tuning time of optical responses of metasurfaces down to sub-picosecond scale has been achieved, which represents the highest tuning speed attainable by far among various mechanisms. The ultrahigh operation speed of optically driven tunable metasurfaces renders them particularly attractive in a number of future applications such as next generation 6G optical communication, LiDAR and photonic neuromorphic computing, to name a few. Despite the rapid development of optically driven tunable metasurfaces in recent years, there is still much room for improvement and remaining challenges have to be addressed before their deployment in practical applications.

For example, a majority of previous works employed active materials (*e.g.*, TCOs and semiconductors) in their bulk forms for generating free-carriers, yet it would be interesting to study similar effects in nanomaterials with further reduced dimensions, *e.g.*, thickness down to a few nm, where quantum confined effects become relevant and might bring forth new tuning mechanisms.[246] A few 2D materials, such as graphene[250] and $MoSe_2$[251], have been demonstrated for constructing tunable metasurfaces by optical pumping induced free-carrier generation. However, the use of 2D materials for developing optically driven tunable metasurfaces is somehow unexplored. In addition, the heterostructures between 2D materials also present opportunities for achieving optical tuning by leveraging their tunable interlayer interactions.[83, 252] For approaches based on the opto-thermal effect, despite the advantages of large refractive index change, fast switching speed and long-term stability of most PCMs, their phase change often occurs at high temperatures. This adds significant power consumption to the tunable system as it requires constant illumination from ultrashort optical pulses. It is worth noting that close-to-room temperature (42˚) phase transition was previously achieved in tungsten-doped VO$_2$.[253] Currently, the use of optical nonlinearity for ultrafast tuning also requires pumping from high-power and ultrashort laser pulses due to the small nonlinear susceptibility of most nonlinear materials. To push it towards commercial usage, it necessitates a substantial reduction of pumping power via either boosting the nonlinear conversion efficiency or lowering the threshold through nano-optical enhancement of the local electric field. A comparison of representative different tuning methods and their modulation depth are summarized in Table 1.

Table 1. A comparison of representative tunable metasurfaces under external stimuli.

| External stimuli | Tunning mechanisms | Active materials | Working spectrum | Applications | Modulation depth/speed |
|---|---|---|---|---|---|
| Electrical stimuli | Free Carrier | ITO | NIR | Modulator [45] | 33% / 826 kHz |
| | | | | Beam steering [44] | 22º / NA |
| | | InAs | MIR | Thermal Modulator [29] | 3.6% / NA |
| | Molecular orientation | LC | Visible | Color filter [51] | NA/NA |
| | | | | Beam steering [35] | 11º / NA |
| | | | | Varifocal lens [60] | NA/1250Hz |
| | | | THz | Modulator [53] | 96% / NA |
| | | | | Beam steering [54] | 32º / NA |
| | Pockels Effect | BaTiO$_3$ | NIR | Modulator [66] | 0.15% / 20MHz |
| | | LN | Visible | Modulator [67] | 82% / 50Hz |
| | Phase transition | GST | Visible | Modulator [73] | >75% / ~2MHz |
| | | VO$_2$ | NIR | Beam steering [74] | 44º / ~ kHz |
| | | | | Modulator [75] | 33% / 787Hz |
| Optical stimuli | Free Carrier | ITO | NIR MIR | Modulator [16] | 40% & 60% / >1000 GHz |
| | | Black phosphorous | NIR / MIR | Modulator [82] | 10%-20% / >1000 GHz |
| | | $PbI_2$ | THz | Modulator [84] | 50% / ~10 GHz |
| | | $Bi_2Se_3$ | THz | Modulator [85] | 31% / ~100 GHz |
| | Phase transition | GST | MIR | Modulator [93] | 81% / NA |
| | | | THz | Modulator [89] | 100% /~500 GHz |
| | | GSST | NIR | Hologram display & optical vortex [98] | NA / NA |
| | | $VO_2$ | THz | Modulator [90] | 50% / > 50 GHz |
| | | $Sb_2S_3$ | Visible | Color display [70] | NA / NA |
| | ENZ effect | CdO:In | MIR | Light source [115] | NA / >1000 GHz |
| Other stimuli | Structural deformation | Au | THz | Half- and Quarter-Waveplates [128] | NA / NA |
| | | $TiO_2$ | Visible | Polarization-Insensitive [135] | NA / NA |
| | | MEMS | NIR | Photodetector [138] | 21º / NA |
| | Chemical reactions | PEDOT:Sulf | MIR | Display [148] | NA / NA |
| | | $Mg-MgH_2$ | Visible | Display [146] | NA / NA |
| | Magnetic Modulation | $N_{18}Fe_{19}$/Au | MIT | Sensing [149] | NA / NA |

In addition to the specific challenges for each individual mechanism, there are also challenges for all optical tuning mechanisms. A major challenge lies in the synergistic integration of high-density large-scale metasurface structures with pixel-wise controllable pumping light beams for independent addressing of individual meta-atoms. This is essential for the realization of tunable metasurface devices that are capable of fully controlling the EM responses of each constituent meta-atoms and it is critical for many real-life applications, such as beam-steering for LiDAR and active holography. Such integration in its current form would impose a high demand on the development of compact and integrated tunable light sources that could provide highly flexible and programable spatiotemporal pumping conditions yet to be achieved. Furthermore, the diffraction-limited size of the pumping beam will impose a fundamental limit on the pixel density of metasurface devices. Alternatively, on-chip integration of light sources with PICs might offer possibilities to overcome the density limitation by exploring the extreme confinement of EM fields enabled by novel plasmonic, photonic and phononic components. On the other hand, for optically driven tunable metasurfaces, the increase in meta-atom density and device area in principle would not sacrifice the operation speed, which, however, poses a speed bottleneck for electrically driven tunable metasurfaces due to the increasing circuit RC constant.

In spite of formidable challenges, the development of optically driven tunable metasurfaces has created both new phenomena and emerging areas. One emerging research field centers around a kind of metasurfaces termed time-varying/spatiotemporal metasurfaces, which features ultrafast temporal modulation of optical properties in a time scale close to the bandwidth of the optical signal.[254] Unlike conventional static metasurfaces which obey fundamental physical laws such as Lorentz reciprocity and conservation of frequency, time-varying metasurfaces could lead to exotic phenomena including frequency conversion and time refraction,[78, 255, 256] non-reciprocal Snell's law for reflection and refraction,[257, 258] and negative refraction[259, 260]. These exotic physical phenomena are only accessible through time-varying metasurfaces with a modulation speed close to or higher than the bandwidth of optical signal (~ GHz to THz), which largely rely on the employment of optical tuning approaches at present. Optically driven tunable metasurfaces could also play important roles in the implementation of photonic neuromorphic computing hardware, which holds great potential in fast, massively parallel and energy efficient information processing for machine learning and artificial intelligence. In particular, all-optical controlled tunable attenuators based on PCMs[261] could be employed in the implementation of weighted optical interconnects (synapses) in photonic neuromorphic computing systems, in which the synapses weight can be varied by optically controlling the transmission properties of on-chip metasurfaces. Compared to electronic interconnects in neuromorphic electronics, optical interconnects excel greatly in speed, bandwidth, energy efficiency and system simplicity.

Another note is that the versatility of those tuning methods is subjected to the operation wavelength of the device. In THz and GHz regimes, the size of pixels is large enough so that those pixels can be individually switched without any interference with neighboring pixels via electrical contact pads; however, at the visible and NIR regimes, this remains as one difficulty due to the fabrication limit at the nm scale. One potential solution could harness well-established through-silicon via (TSV) technology and 3D integration process to develop dedicated processes to separate meta-atoms/pixels at two different layers to provide electrical connection and achieve almost zero interference simultaneously.

**Advanced design methodologies** to achieve unprecedented performance is indispensable to the translation of the disruptive concept of tunable/reconfigurable metasurfaces to on-demand and adaptive nanophotonics. The rapid developing ML algorithms and AI-driven approaches have been considered as a powerful engine for intelligent analysis and optimization of a big amount of data of meta-atoms. With the ever-increasing evolution of ML and its branch, deep learning, and their incredible capacity provide accurate decision making in the fields of traffic prediction, image and speech recognition, self-driving cars, *etc*. we are now witnessing their applicability to address challenges of dynamic control of EM waves. Due to its potential to offer high accuracy, fast speed, less demanding on computational resources, and flexibility, ML has opened up exciting opportunities for the design of meta-atoms with arbitrary and complex geometry and even unknown materials properties, which is hardly achievable with conventional design methods. AI-assisted structural and functional design methodologies could help us better understand the underpinning physics of tunable metasurface and could accelerate the realization of next generation tunable metasurface.

# AUTHOR INFORMATION

**Corresponding Authors**

**Xiao Renshaw Wang** — School of Physical and Mathematical Sciences, Nanyang Technological University, Singapore 637371, Singapore; School of Electrical and Electronic Engineering, 50 Nanyang Avenue, Nanyang Technological University, Singapore 639798, Singapore.  orcid.org/0000-0002-5503-9899; Email: renshaw@ntu.edu.sg

**Qi Jie Wang** — School of Physical and Mathematical Sciences, Nanyang Technological University, Singapore 637371, Singapore; School of Electrical and Electronic Engineering, 50 Nanyang Avenue, Nanyang Technological University, Singapore 639798, Singapore.  orcid.org/0000-0002-9910-1455; Email: qjwang@ntu.edu.sg

**Hong Liu** — Institute of Materials Research and Engineering, Agency for Science, Technology and Research (A*STAR), 2 Fusionopolis Way, Singapore 138634, Singapore  orcid.org/0000-0002-3560-9401; Email: h-liu@imre.a-star.edu.sg

## Authors

**Omar A. M. Abdelraouf** — School of Physical and Mathematical Sciences, Nanyang Technological University, Singapore 637371, Singapore;

**Ziyu Wang** — Institute of Materials Research and Engineering, Agency for Science, Technology and Research (A*STAR), 2 Fusionopolis Way, Singapore 138634, Singapore;  orcid.org/0000-0002-0959-4787

**Hailong Liu** — Institute of Materials Research and Engineering, Agency for Science, Technology and Research (A*STAR), 2 Fusionopolis Way, Singapore 138634, Singapore;  orcid.org/0000-0001-9853-9327

**Zhaogang Dong** — Institute of Materials Research and Engineering, Agency for Science, Technology and Research (A*STAR), 2 Fusionopolis Way, Singapore 138634, Singapore;  orcid.org/0000-0002-0929-7723

**Qian Wang** — Institute of Materials Research and Engineering, Agency for Science, Technology and Research (A*STAR), 2 Fusionopolis Way, Singapore 138634, Singapore;  orcid.org/0000-0003-0750-7843

**Ming Ye** — School of Electrical and Electronic Engineering, 50 Nanyang Avenue, Nanyang Technological University, Singapore 639798, Singapore.  orcid.org/ 0000-0002-5961-7991

## Author Contributions

## Notes

The authors declare no competing financial interest.

# ACKNOWLEDGMENTS

The authors would like to acknowledge the funding support from Singapore Ministry of Education Academic Research Fund Tier 2 under grant no. MOE2018-T2-1-176, MOE-T2EP50220-0016 and by A*STAR AME programmatic grant (grant no. A18A7b0058), IMRE project (SC25/18-8R1804-PRJ8), AME IRG grant (Project No. A20E5c0094 and A20E5c0095).